\documentclass[aps,prc,superscriptaddress,showpacs,twocolumn,floatfix]{revtex4}

\usepackage{graphicx}
\usepackage{amsmath}
\usepackage{longtable}

\def\gtorder{\mathrel{\raise.3ex\hbox{$>$}\mkern-14mu
 \lower0.6ex\hbox{$\sim$}}}
\def\ltorder{\mathrel{\raise.3ex\hbox{$<$}\mkern-14mu
 \lower0.6ex\hbox{$\sim$}}}
\def\gep{G_E^p}
\def\gmp{G_M^p}
\def\gegm{G_E^p / G_M^p}
\def\gen{G_E^n}
\def\gmn{G_M^n}

\def\etal{\emph{et al.}}

\begin{document}

\title{Flavor decomposition of the nucleon electromagnetic form factors}

\author{I.~A.~Qattan}
\affiliation{Khalifa University of Science, Technology and Research, Department
of Applied Mathematics and Sciences, P.O. Box 573, Sharjah, U.A.E}

\author{J.~Arrington}
\affiliation{Physics Division, Argonne National Laboratory, Argonne, Illinois,
60439, USA}

\date{\today} 

\begin{abstract}

\textbf{Background:}  The spatial distribution of charge and magnetization
in the proton and neutron are encoded in the nucleon electromagnetic form
factors.  The form factors are all approximated by a simple dipole function,
normalized to the charge or magnetic moment of the nucleon.  The
differences between the proton and neutron form factors and the deviation
of $\gen$ from zero are sensitive to the difference between up- and down-quark
contributions to the form factors.

\textbf{Purpose:} Recent measurements of $\gen$ up to 3.4~(GeV/c)$^2$ allow
for a much more detailed examination of the form factors.  The
flavor-separated form factors provide information on the quark flavor
dependence of the nucleon structure and test theoretical models of the form
factors.

\textbf{Methods:} We combine recent measurements of the neutron form factors
with updated extractions of the proton form factors, accounting for two-photon
exchange corrections and including an estimate of the uncertainties for all of
the form factors to obtain a complete set of measurements up to $Q^2 \approx
4$~(GeV/c)$^2$.  We use this to extract the up- and down-quark
contributions which we compare to recent fits and calculations.

\textbf{Results:} We find a large differences between the up- and down-quark
contributions to $G_E$ and $G_M$, implying significant flavor dependence in
the charge and magnetization distributions.  The rapid falloff of the ratio
$\gep/\gmp$ does not appear in the individual quark form factors, but arises
from a cancellation between the up- and down-quark contributions. We see
indications that the down-quark contributions to the Dirac and Pauli form
factors deviate from the suggested 1/$Q^4$ scaling behavior suggested by
a previous analysis. While recent models provide a generally good
qualitative description of the data, the down-quark contribution to $G_E/G_M$
and $F_2/F_1$ are not reproduced by any of the models.  Finally, we note
that while the inclusion of recent $\gmn$ data from CLAS modifies the
high-$Q^2$ slightly, the tension between these data and previous measurements
at lower $Q^2$ has a more significant impact, suggesting the need for 
additional data in this region.

\end{abstract}


\maketitle

\section{Introduction}

The nucleon electromagnetic form factors provide information
on the spatial distributions of charge and magnetization of the
nucleon~\cite{sachs62}, corresponding to a Fourier transform of the
nucleon's charge or magnetization density in a non-relativistic picture.
As such, the form factors provide some of the most direct constraints on 
the partonic structure of the nucleon.  The form factors are functions
only of the four-momentum transfer squared, $Q^2$, and can also be expressed
in terms of the Dirac, $F_1(Q^2)$, and Pauli, $F_2(Q^2)$, form factors,
which are related to the electric and magnetic form factors:
\begin{eqnarray} \label{eq:protonElecMagFFs}
G_E(Q^2) = F_1(Q^2) - \tau F_2(Q^2), \nonumber \\
G_M(Q^2) = F_1(Q^2) + F_2(Q^2),
\end{eqnarray}
where $\tau = Q^2/4M_N^2$ and $M_N$ is the mass of the nucleon.
In the limit $Q^2 \to 0$, $G_E$ and $G_M$ become the charge and magnetic
moment of the nucleon, while $F_1$ and $F_2$ yield the charge and
anomalous magnetic moment.

It has long been known that $\gep$, $\gmp$, and $\gmn$ approximately
follow the dipole form, $G_D = (1+Q^2/Q_0^2)^{-2}$ with $Q_0^2 =
0.71$~(GeV/c)$^2$, up to $Q^2$=5-10~(GeV/c)$^2$, while the neutron electric
form factor, $\gen$, is close to zero. This observation is consistent with the
simple, non-relativistic interpretation in which the charge and magnetization
of the nucleon is carried by the quarks, and the up and the down quark have
similar spatial distributions.  This yields identical contributions for all
form factors except for $\gen$, for which there is a nearly complete
cancellation between the up- and down-quark charge distributions.  As such,
measurements of $\gen$ played an important role, demonstrating that there is a
measurable difference between the up- and down-quark distributions.  This
is qualitatively consistent with the pion cloud picture of the neutron, where
a positive core and negative cloud arise from a virtual p+$\pi^-$ component of
the neutron structure~\cite{fermi47, rosenbluth50, friedrich03, crawford10,
gentile11}.

Details of the recent progress in measurements of the nucleon electromagnetic
form factors can be found in recent global analyses and
reviews~\cite{arrington07, venkat11, arrington07a, perdrisat07, arrington11a,
CJRW2011}. Several of these new measurements demonstrate the limitations of
the simple, nonrelativistic picture. The decrease of the ratio $\gegm$ with
$Q^2$, as observed in polarization measurements~\cite{punjabi05, gayou01,
puckett12, puckett10}, provided a clear demonstration that the form factors
were not simply the sum of dipole-like contributions from the up and down
quarks. While these results were inconsistent~\cite{arrington03a, arrington04a,
qattan05} with earlier extractions based on Rosenbluth separation
techniques~\cite{rosenbluth50}, this is now widely believed to be the result
of small two-photon exchange (TPE) corrections which yield a small angular
dependence to the cross section~\cite{carlson07, arrington11b}, mimicking the
small signal expected from the contribution of $\gep$ at high $Q^2$. More
recent measurements of the neutron form factors~\cite{lachniet09, riordan10}
have provided a complete set of data up to $Q^2$=3.4~(GeV/c)$^2$, which has
enabled a detailed comparison of the up- and down-quark contribution from the
high-$Q^2$ $\gen$ measurements, as well as a detailed comparison of the proton
and neutron magnetic form factors.

Assuming isospin and charge symmetry and neglecting the contribution of
strange quarks allows us to express the nucleon form factors in terms of the
up- and down-quark contributions~\cite{miller90, beck01},
\begin{eqnarray}
G_{E,M}^p = \frac{2}{3} G_{E,M}^u - \frac{1}{3} G_{E,M}^d, \nonumber \\
G_{E,M}^n = \frac{2}{3} G_{E,M}^d - \frac{1}{3} G_{E,M}^u.
\label{eq:flavorsep}
\end{eqnarray}
This yields the following expression
for the up- and down-quark contributions to the proton form factors:
\begin{eqnarray}
G_{E,M}^u = 2 G_{E,M}^p + G_{E,M}^n,~~ 
G_{E,M}^d = G_{E,M}^p + 2 G_{E,M}^n,
\end{eqnarray}
with similar expressions for $F_1$ and $F_2$.  In this convention,
$G_{E,M}^u$ represents the up-quark distribution in the proton and
the down-quark distribution in the neutron. Because the charge is factored out
from the up- and down-quark contributions, the $Q^2=0$ values for these form
factors are $G_E^u=2$, $G_E^d=1$, while the quark magnetic moments are taken
to be the $Q^2=0$ limit of the magnetic form factors: $\mu_u = (2\mu_p+\mu_n)
= 3.67 \mu_N$ and $\mu_d=(\mu_p+2\mu_n) = -1.03 \mu_N$.
Note that the up- and down-quark contributions as
defined here are the combined quark and antiquark contributions, and so
represent the difference between the quark and antiquark distributions, due to
the charge weighting of the quark and antiquark contributions to the form
factors.

\section{Form factor input and two-photon exchange corrections} \label{formfactor_tpe}

Recently, the ratio $R_n = \mu_n\gen/\gmn$ of the neutron was measured at
Jefferson Lab up to $Q^2=$3.4~(GeV/c)$^2$~\cite{riordan10}.  These data,
combined with $R_p = \mu_p\gep/\gmp$ measurements in the same $Q^2$
range~\cite{punjabi05, gayou01, puckett12} allowed for the first time a
comparison of the behavior of $F^n_2/F^n_1$ and $F^p_2/F^p_1$, as well as a
separation of the up- and down-quark contributions to the form factors. In the
pioneering work of~\cite{CJRW2011}, which will be referred to as ``CJRW''
throughout this text, measurements of $R_n = \mu_n\gen/\gmn$ for
$0.30<Q^2<3.40$ (GeV/c)$^2$ were combined with parameterizations of $\gep$,
$\gmp$, and $\gmn$~\cite{kelly04} to examine the flavor-separated
contributions and the ratio
\begin{equation} \label{eq:F2F1pnRatio}
\frac{F^{(p,n)}_2}{F^{(p,n)}_1} = \Big(\frac{1-R_{(p,n)}/\mu_{(p,n)}}
{\tau+R_{(p,n)}/\mu_{(p,n)}} \Big).
\end{equation}
In the CJRW analysis, only the uncertainty from $R_n$ was included
in the analysis, as this was the largest source of uncertainty in the
quantities they examined.  Thus, the results of the CJRW analysis for any
quantities which did not depend on $\gen$ as shown here will simply reflect
the parameterizations of the other form factors and have no associated
uncertainties.

Refs.~\cite{riordan10, CJRW2011} provided the first results for
flavor-separated form factors at high $Q^2$ values, and demonstrated a
significant difference between the up- and down-quark contributions.
We expand on their analysis mainly be accounting for two effects that were not
included in their initial result. First, we include uncertainties associated
with all of the form factors, as they are an important contribution for some of
the extracted quantities. In addition, the proton form factor
parameterization~\cite{kelly04} used in their analysis did not apply any
two-photon exchange corrections, although data was selected with an eye
towards reducing the impact of TPE corrections.  We address this by using an
extraction of the proton form factors which includes TPE corrections.

For $\gmn$, the CJRW results also used the parameterization of
Kelly~\cite{kelly04}.  However, recent data from the CLAS
collaboration~\cite{lachniet09} shows smaller deviations from the dipole
form at high $Q^2$, and thus will have a small impact on the high-$Q^2$
behavior of the up- and down-quark contributions to the magnetic form factor.
We use an updated parameterization to world's $\gmn$ data~\cite{lung93,
anklin94, anklin98, kubon02, anderson06, lachniet09} using the
same form as Kelly, but obtaining modified parameters: $a_1 = 5.857$,
$b_1 = 18.74$, $b_2 = 54.07$, and $b_3 = 177.73$.  We take the uncertainty to
be the same as in the original Kelly fit, using the full error correlation
matrix~\cite{plaster_private}.  This error band is fairly consistent with the
experimental uncertainties with the new CLAS data included, as the simple
functional form yielded a very small uncertainty in the Kelly analysis for
regions where there were limited data. The updated fit yields a small
modification to the high-$Q^2$ behavior of $\gmn$, but also reduces the value
of $\gmn$ for $Q^2$ values near 1--1.5~(GeV/c)$^2$. The updated parameterization
falls in between the earlier data below 1~(GeV/c)$^2$~\cite{anklin98, kubon02} and
the new CLAS data above $Q^2=1$~(GeV/c)$^2$~\cite{lachniet09}.  Where the updated
fit has a significant impact, we will compare the results obtained using the
Kelly fit and our updated parameterization.

For the neutron electric form factor, we take the fit to $R_n =
\mu_n\gen/\gmn$ from Riordan~\etal~\cite{riordan10}. Taking the full error
correlation matrix for $R_n$ and $\gmn$~\cite{riordan_private}
yields uncertainties on $\gen$ that are significantly smaller than the
uncertainties on the individual measurements, due to the simple functional form
of the $R_n$ parameterization.  To account for this, we scale up the uncertainty
on $\gen$ by a factor of two to provide more realistic uncertainties in the
flavor-separated results.  Thus, any quantities which do not depend on 
the proton form factors (e.g. the up- and down-quark contributions to
the magnetic form factor) will simply reflect the above parameterizations.
This is similar to the CJRW analysis for quantities which do not include the
$\gen$ measurements, although we include a realistic estimate of the
uncertainties in the parameterized form factors, while their results that
do not depend on $\gen$ will not show any uncertainty.

The leading TPE effect on the electron-proton elastic scattering cross
section, $\sigma_{ep}$, comes from the interference of the one- and two-photon
exchange amplitudes which yield a small correction to both the cross section
and recoil polarization measurements. The angular dependence of this
correction to the cross section can include a much larger effect on the
extracted form factors~\cite{guichon03}, while the recoil polarization data do
not have a similar amplification of the effect.  Recent measurements of the
angular dependence of $R_p$ also suggest small TPE contributions to the
polarization measurements~\cite{meziane2011}.  To account for the TPE
contribution to $\sigma_{ep}$, one can add an additional term which forces the
Rosenbluth extraction to yield the same value of $\gep/\gmp$ as the
polarization transfer data.  We account for TPE effects by using the
extraction of $\gep$ and $\gmp$ from an analysis which constrains the TPE
corrections based on the discrepancy between Rosenbluth and polarization
measurements~\cite{qattan2011b}.  We use the form factors extracted based on
the TPE parametrization from Borisyuk and Kobushkin (BK
parametrization)~\cite{borisyuk11}, which takes the corrections to be
linear~\cite{tvaskis06} in the virtual photon polarization parameter,
$\varepsilon$, and constrains the correction to vanish in the limit of small
angle scattering, as expected from charge conjugation and crossing
symmetry~\cite{chen07, arrington11b}, and as observed in comparisons of
positron-electron scattering~\cite{arrington04b}.

In this analysis, we take the extraction of $\gep$ and $\gmp$ from
Ref.~\cite{qattan2011b} for several electron-proton scattering
measurements~\cite{andivahis94, walker94, christy04, qattan05, bartel73,
litt70, berger71}, and add data at lower $Q^2$ from Ref.~\cite{janssens65}
analyzed following the same procedure. This provides values of $\gep$ and
$\gmp$,with TPE corrections constrained by polarization transfer data.  In the
analysis of Ref.~\cite{qattan2011b}, no uncertainty is applied for the
parameterization of $R_p$. For our analysis, we include an additional
uncertainty of $\delta_{R}=0.01 Q^2$ ($Q^2$ in (GeV/c)$^2$) in the
polarization ratio $R_p$.

This approach to constraining TPE is not expected to be as reliable at low
$Q^2$ values, as the difference between the two techniques is significantly
smaller in this region, and because the parameterization of $R_p$ from 
polarization data included very little low-$Q^2$ data.  However, recent
low-$Q^2$ measurements~\cite{crawford07, ron07, zhan11, ron11} suggest that the
parameterization is relatively reliable for the range of data examined here,
and the agreement between these polarization measurements and new
Rosenbluth separation data~\cite{bernauer10} support the idea that the
corrections are relatively small.  Thus, the final result should be somewhat
insensitive to the exact details of the extraction of the TPE effects in this
region.  Because the TPE corrections are still important in the limit of
low $Q^2$~\cite{arrington11b, blunden05b, bernauer11, arrington11c},
we compare our results to an extraction of the flavor-separated form factors
using the proton form factor parameterization from Refs.~\cite{arrington07}
and~\cite{venkat11}.  Both of these extractions include TPE corrections
calculated in a hadronic framework~\cite{blunden03, blunden05a} which is
expected to be more reliable at low values of $Q^2$, and is in good agreement
with other low $Q^2$ calculations~\cite{borisyuk06, borisyuk07a, borisyuk08, 
borisyuk12} (as shown in Ref.~\cite{arrington12c}).  Comparison to these 
fit allows for a check of our low-$Q^2$ phenomenological TPE
extraction.  The fit of Venkat~\etal~\cite{venkat11} also includes additional
polarization data, in particular at low $Q^2$ values~\cite{crawford07, ron07,
zhan11, ron11}, and includes a more careful evaluation of the low-$Q^2$
behavior of the fit.

Note that a more complete flavor separation at the lowest $Q^2$ values would
involve an updated extraction of the form factors, including calculated TPE
corrections and the most recent form factor data~\cite{crawford07, ron07,
zhan11, ron11, bernauer10}, along with constraints on strange-quark
contributions taking measurements of elastic parity-violating electron
scattering~\cite{spayde04, maas04, aniol04, maas05a, armstrong05, young06,
acha07, liu07, pate08, wang09, androic10, paschke11, ahmed12, armstrong12}.
For this work, the primary focus is at somewhat higher $Q^2$ data, and the
comparison of our TPE corrections~\cite{qattan2011b} to the hadronic
corrections~\cite{blunden05a} applied in the recent proton
fits~\cite{arrington07, venkat11} should provide an idea of the robustness of
the low $Q^2$ results.

\section{Recent Theoretical Predictions} \label{recent_theory} 

In this section we summarize several recent theoretical studies of
the nucleon elastic form factors which we will compare to our extracted
flavor-separated form factors.

Clo{\"e}t~\etal~\cite{cloet09} presented a calculation of a
dressed-quark core contribution to the nucleon electromagnetic form factors
defined by the solution of a Pioncar\'{e} covariant Faddeev equation. 
This calculation includes dressed-quark anomalous magnetic moment
within the framework of Dyson-Schwinger equations (DSEs). The Faddeev equation
was described by specifying that quarks are dressed, and two of the three
dressed quarks are always correlated as color-\=3 diquarks. The nucleon is
represented by a Faddeev amplitude of the form $\Psi=(\Psi_1+\Psi_2+\Psi_3)$
with $\Psi_{1,2}$ obtained from $\Psi_3$ by a cyclic permutation with
$\Psi_3(p_i,\alpha_i,\tau_i)$ expressed as sum of
scalar- and axial-vector-diquark correlations with $(p_i,\alpha_i,\tau_i)$
being the momentum, spin, and isospin labels of the quarks. The Faddeev
equation satisfied by $\Psi_3$ was constructed by specifying the dressed-quark
propagator, diquark Bethe-Salpeter amplitudes, and the diquark propagators.
The nucleon-photon vertex was calculated using six diagrams, with photon
coupling to the quark or the diquark along with loop and exchange terms. In
this approach, the Faddeev equation has only two new parameters in the
nucleon sector: the masses of the scalar and axial-vector diquarks. The scalar
mass is set by requiring a nucleon mass $M_N=$ 1.18 GeV, and the axial-vector
mass is chosen so as $M_\Delta=$ 1.33 GeV. The proton has a mass larger than
the physical value to allow for additional contributions from the
pseudoscalar mesons. The quark, diquark, and exchange (two body) contributions
to the nucleons form factors were calculated up to $Q^2$=12~(GeV/c)$^2$. In
addition, the decomposition according to diquark spin and flavor
contributions were also calculated for the same $Q^2$ range. The DSE
approach aims to simultaneously describe meson and baryon
physics~\cite{roberts07} with only a few parameters, most of which are fixed
to static properties of the pion. This calculation will be referred to as
``DSE'' throughout this text. Because the calculation does not include pion
cloud contributions, the masses and magnetic moments are not expected to
reproduce the physical values.  In Ref.~\cite{cloet09}, comparisons to data
were made in terms of $y=Q^2/M^2$. In our analysis, we evaluate their
parameterization in terms of $y$ using the physical nucleon mass, and the
experimental magnetic moments of the proton and neutron. Rescaling to the
physics magnetic moments removes the discrepancies at $Q^2=0$, although to the
extent that the calculation leaves room for additional pion contributions,
more important at low $Q^2$, this would be expected to worsen the agreement
at larger $Q^2$ values.

Clo{\"e}t and Miller~\cite{cloet2012} proposed a relativistic constituent
quark model which is constrained by the nucleon form factors but also
reproduces the quark spin content of the nucleon. This is an extension of a
previous light-front calculation which included three constituent
quarks~\cite{frank96} and predicted the falloff of $\gep/\gmp$. In this model,
it is assumed that the quarks are moving in a cloud of pions. The valance
quarks are represented by quark-diquark combination and treated in a way
consistent with Pioncar\'{e} invariance. Due to the long range nature of the
quarks' interactions as mediated by a single pion exchange, a pion emitted by a
nucleon can also be absorbed by the same nucleon, allowing for a pion cloud
contribution. The light-front wave function that describes the interaction of
a quark and diquark to form a nucleon is used to construct the Fock state
needed to represent the nucleon. The quark-diquark approximation includes both
scalar and axial-vector correlations, and the flavor couplings were added to
obtain a symmetric spin-flavor wavefunction. The pion component is introduced
using a single pion loop around the bare nucleon, including diagrams
with the photon coupling to the bare nucleon and coupling to the
nucleon or pion in the pion loop. Terms involving $\gamma N \to \pi N$
couplings and the effects of intermediate $\Delta$ were not included. The
model is finally expressed in terms of ten parameters, representing the quark
and diquark masses and contributions to the light-front wave function and a
parameter describing the high-momentum behavior of the pion-nucleon vertex
function. The parameters were adjusted to provide the best fit to the nucleon
form factors~\cite{kelly04} up to $Q^2=10$~(GeV/c)$^2$. The model also yields
a quark contribution to the proton spin which is found to be in agreement with
experimental evaluations. This calculation will be referred to as pion-cloud
relativistic constituent quark model, ``PC-RCQM''.

Gonzalez-Hernandez~\etal~\cite{hernandez2012} provided an interpretation of
the flavor dependence of the nucleon form factors in terms of Generalized
Parton Distributions (GPDs). They incorporated the Regge contribution into
GPDs that already apply diquark models by introducing a spectral distribution
for the spectator diquark mass $\rho_R(M^2_X)$. Inclusion of the Regge
contributions is crucial to obtain the correct normalized structure functions.
The model can be summarized in the expression $F(X,\zeta,t) = N
G^{M_\Lambda}_{M_X,m}(X,\zeta,t)R_p^{\alpha,\alpha'}(X,\zeta,t)$ where the
flavor dependence of the nucleon form factors was attributed mainly to Reggeon
exchange contributions and handbag (quark-diquark) contribution.
The diquark contribution was later represented by the two functions
$H$ and $E$ and the proton-quark-diquark vertex was parameterized using a
dipole type coupling with two fit parameters $m_q$ and $M_q$.
The final fit to DIS structure functions, nucleon form factors, and deep
virtual Compton scattering were performed, one without the new CJRW
extractions~\cite{CJRW2011}, and one including these results to produce 
improved constraints on the flavor dependence of the GPDs. The flavor separated
nucleon form factors, separated into Regge and diquark contributions, were
calculated up to $Q^2$=5~(GeV/c)$^2$. For $F^u_1$, the diquark contribution
dominates the Regge contribution at low $Q^2$ and both contributions become
comparable at high $Q^2$. On the other hand, the diquark and Regge
contributions are comparable for $F^d_1$ and the Regge contribution dominates
$F^u_2$ and $F^d_2$, in particular at high $Q^2$. This calculation will be
referred to as ``GPD'' throughout this text.

Recently, Rohrmoser~\etal~\cite{rohrmoser11} analyzed the flavor
decomposition of the nucleon electromagnetic form factors within the framework
of a relativistic constituent quark model whose hyperfine interaction is
derived from Goldstone-boson exchange~\cite{glozman98} as a result of
spontaneous breaking of chiral symmetry in low-energy QCD. In this model,
nucleons are represented by three-quark-configuration driven by an interaction
Lagrangian formed by coupling of Goldstone bosons with valance-quark field.
The nucleon wavefunction has no diquark configuration or mesonic effects and
contains non-vanishing orbital angular momenta and mixed-symmetric spatial
wave-function component with relatively small non-symmetric contribution. The
key ingredient of the nucleon wave functions is the interaction of the mass
operator, which has a linear confinement, with the QCD string tension and
chiral symmetry breaking hyperfine interaction.  We show here the latest
theoretical results~\cite{plessas_private} and not the originally
published~\cite{rohrmoser11} and note that the two results differ slightly.
The form factors and their flavor decomposition were calculated up
to $Q^2$=4~(GeV/c)$^2$. Throughout this text this calculation will be referred
to as ``GBE-RCQM''.

Note that the GBE-RCQM and DSE calculations are predictions for the
flavor-separated form factors, as they do not adjust parameters to match the
proton or neutron form factors, while the PC-RCQM and GPD curves are fits to
the data using parameterizations based on the model. The PC-RCQM model has a
total of 10 parameters, while the GPD fit has 16 parameters for each of the
GPDs, which are fit to reproduce both the form factors and parton
distributions, with additional constraints from lattice QCD.  Thus, one
expects these models to better reproduce the data, although this does not
represent as conclusive of a test as it does for the DSE and GBE-RCQM
predictions.  The models which are adjusted to reproduce the form factor data
must also be compared to other measurements, as mentioned above and discussed
in more detail in the original works~\cite{cloet2012, hernandez2012}.


\begin{figure}[!htbp]
\begin{center}
\includegraphics*[width=7.9cm,height=5.0cm]{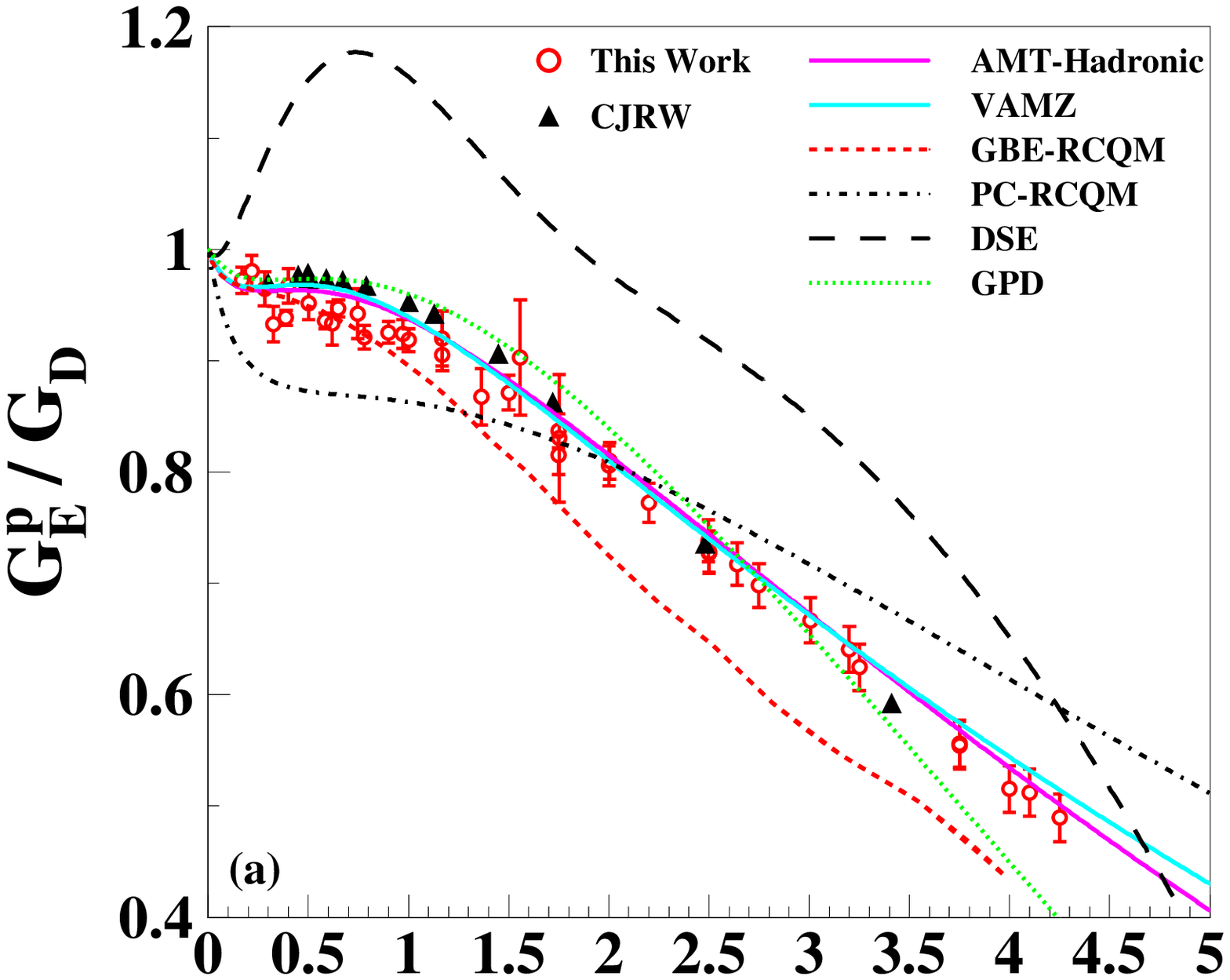}\\
\includegraphics*[width=7.9cm,height=5.0cm]{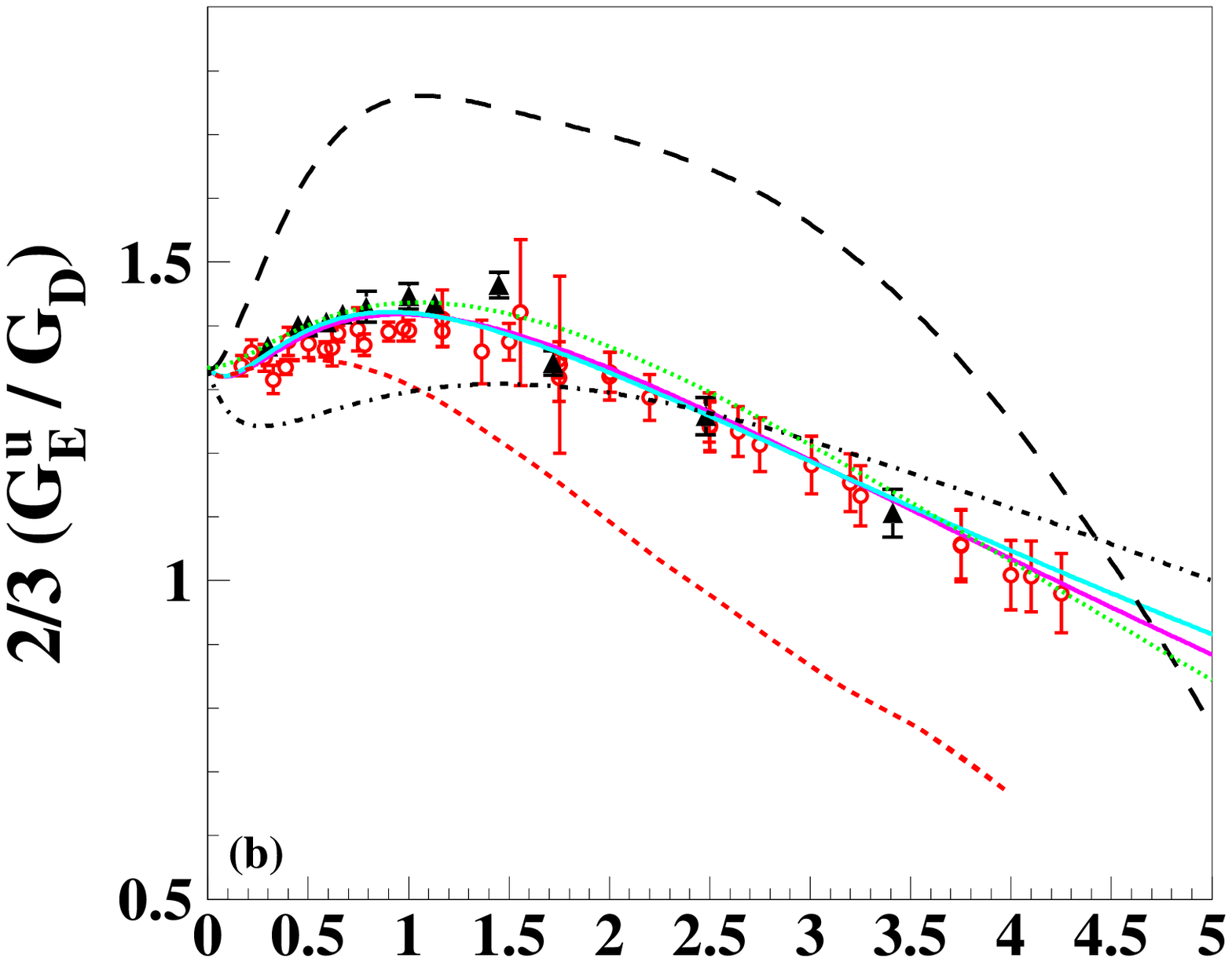}\\
\includegraphics*[width=7.9cm,height=5.5cm]{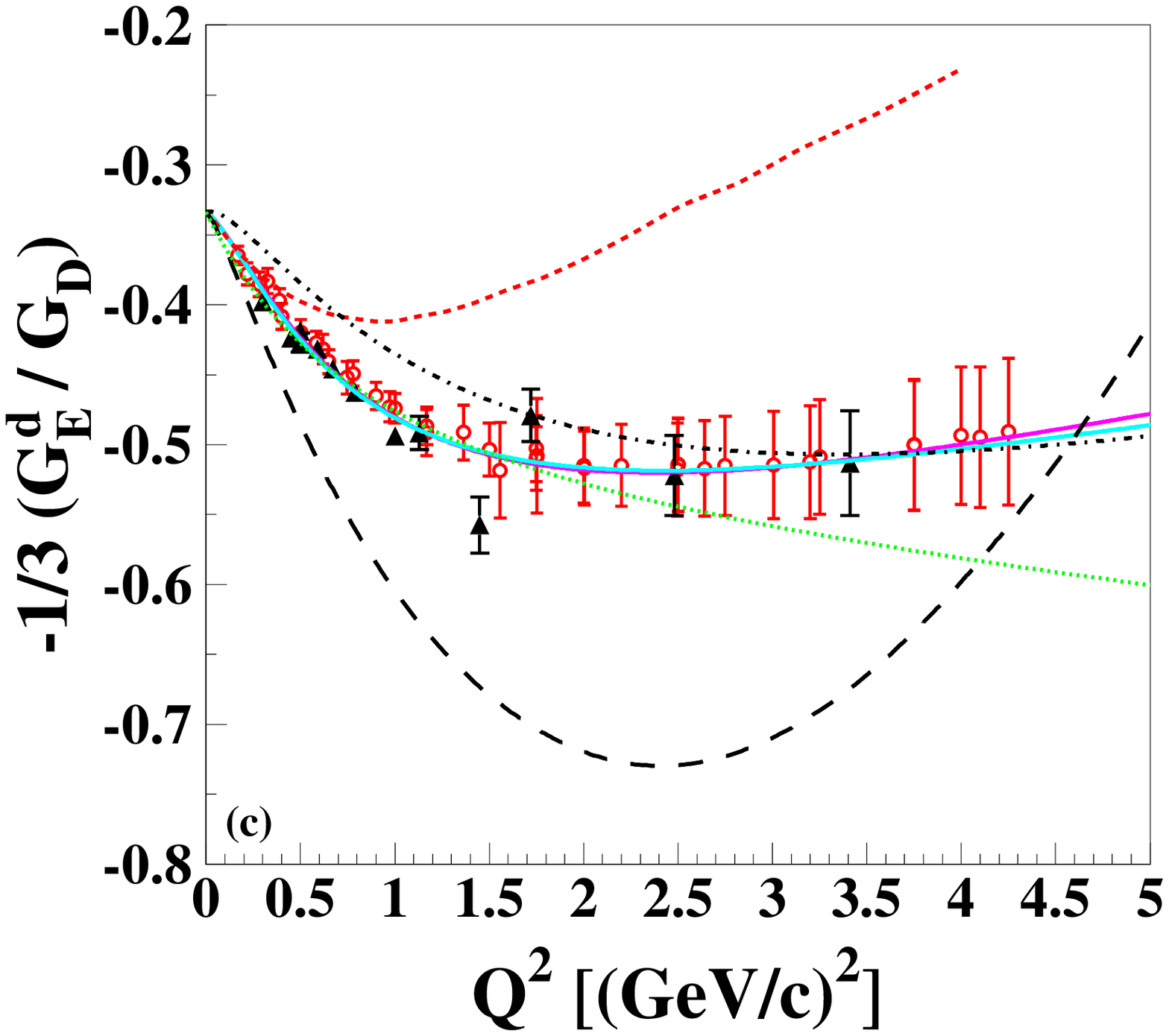}
\end{center}
\vspace{-0.5cm}
\caption{(color online) $\gep/G_D$ [top] and its flavor-separated
contributions $2/3(G_E^u/G_D)$ [middle], and -$1/3(G_E^d/G_D)$ [bottom]
as obtained from Refs.~\cite{andivahis94, walker94, christy04, qattan05,
bartel73, litt70, berger71, janssens65} based on our fit and the CJRW
extractions~\cite{CJRW2011}. Also shown are the AMT~\cite{arrington07} and
VAMZ fits~\cite{venkat11}, and the values from the
GBE-RCQM~\cite{rohrmoser11,plessas_private}, PC-RCQM~\cite{cloet2012}, the
DSE~\cite{cloet09}, and the GPD~\cite{hernandez2012} models.}
\label{fig:GEpFlavor}
\end{figure}

\begin{figure}[!htbp]
\begin{center}
\includegraphics*[width=7.9cm,height=5.0cm]{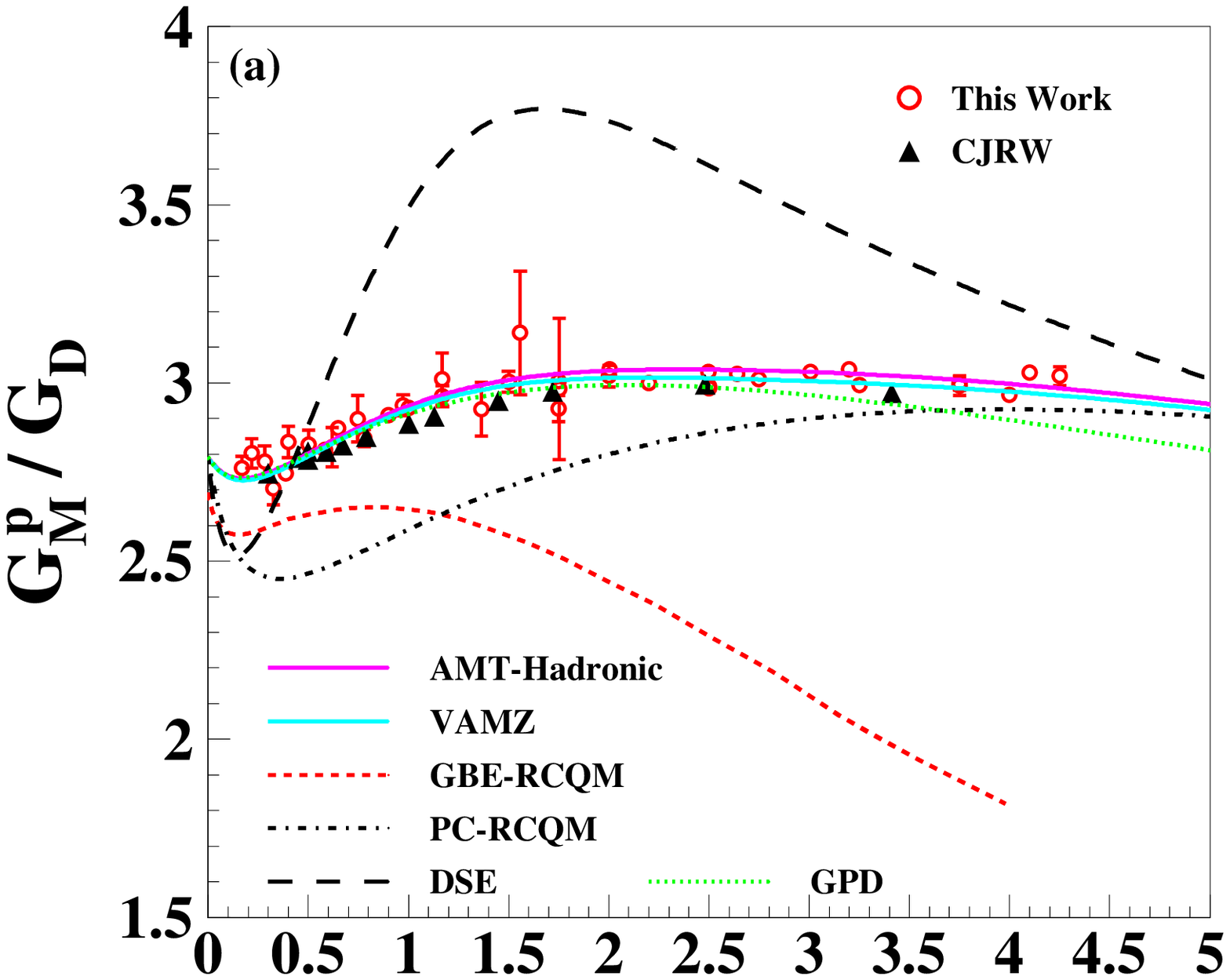}\\
\includegraphics*[width=7.9cm,height=5.0cm]{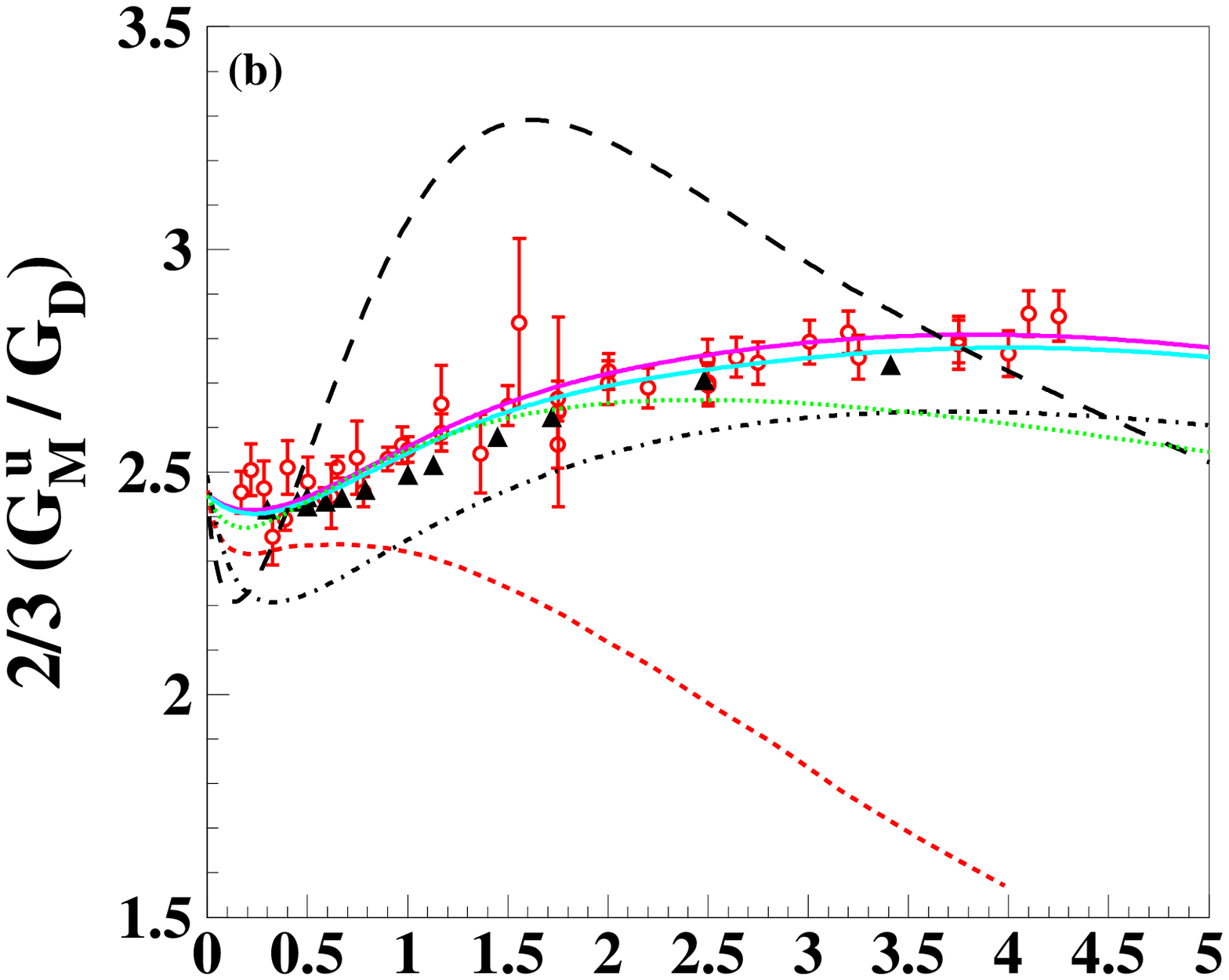}\\
\includegraphics*[width=7.9cm,height=5.5cm]{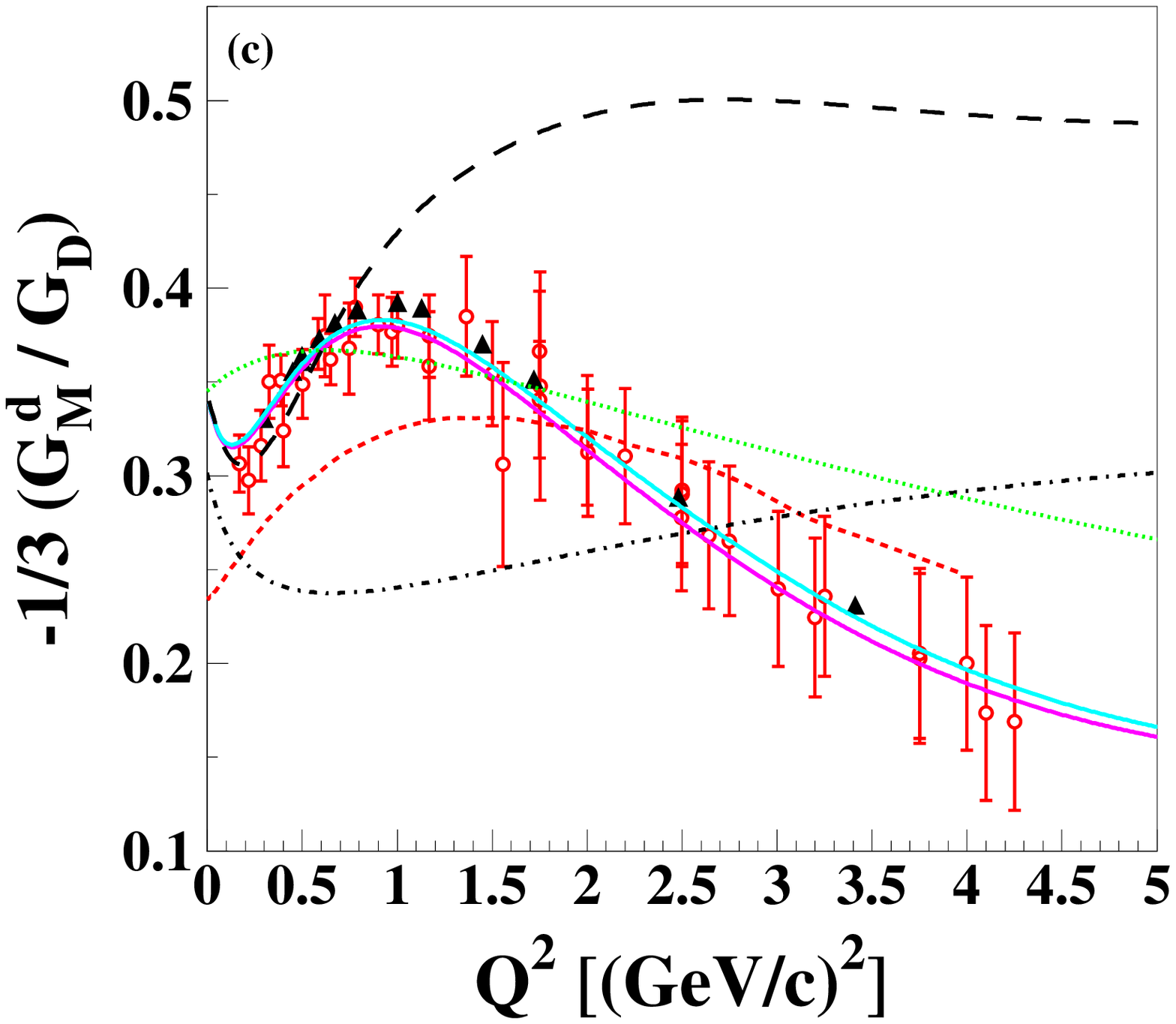}
\end{center}
\vspace{-0.5cm}
\caption{(color online) $\gmp/G_D$ and its flavor-separated contributions.
Curves and data points are the same as in Fig.~\ref{fig:GEpFlavor}.}
\label{fig:GMpFlavor}
\end{figure}

\section{Results and Discussion} \label{result_discussion}

\begin{figure}[!htbp]
\begin{center}
\includegraphics*[width=7.9cm,height=5.0cm]{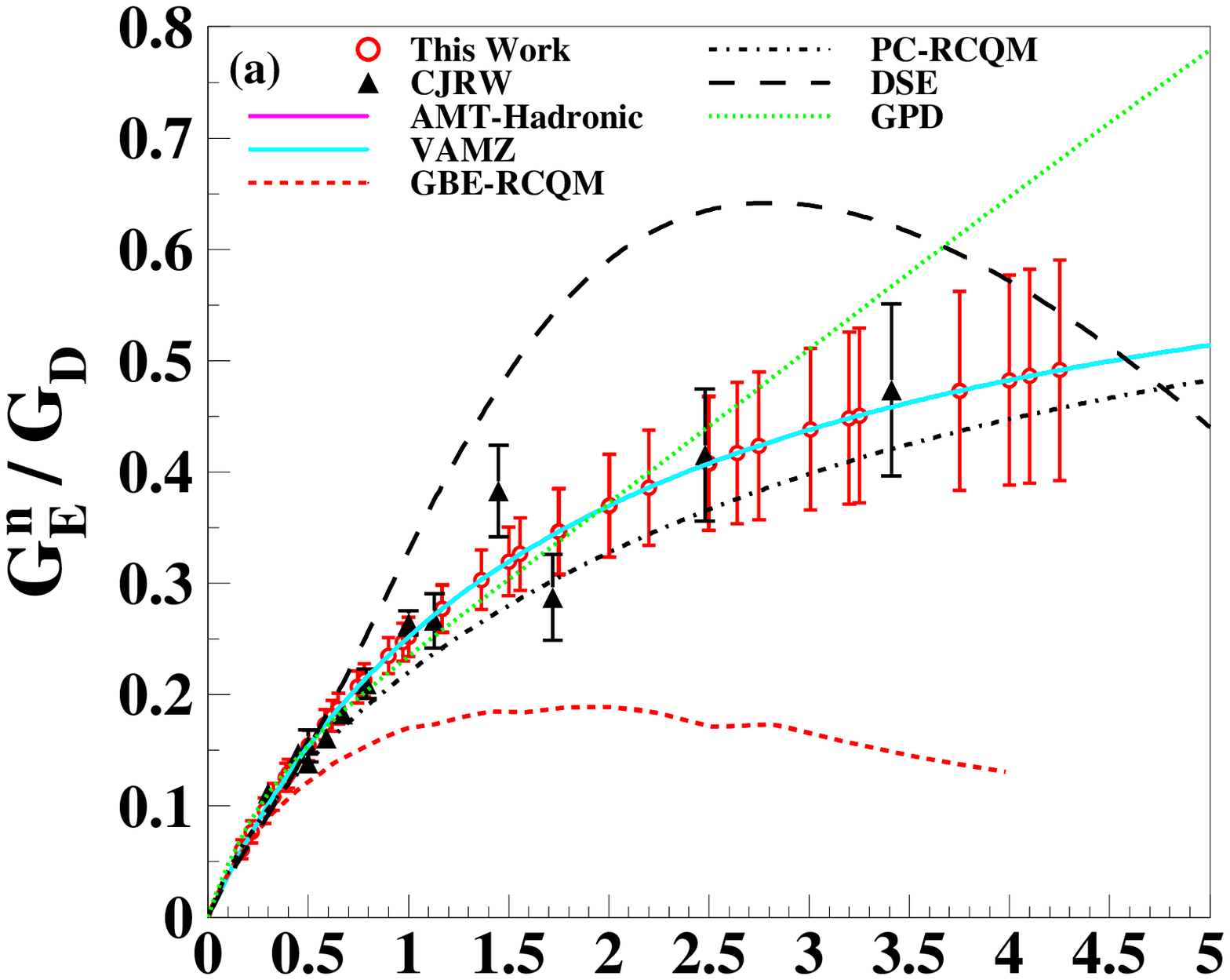}\\
\includegraphics*[width=7.9cm,height=5.0cm]{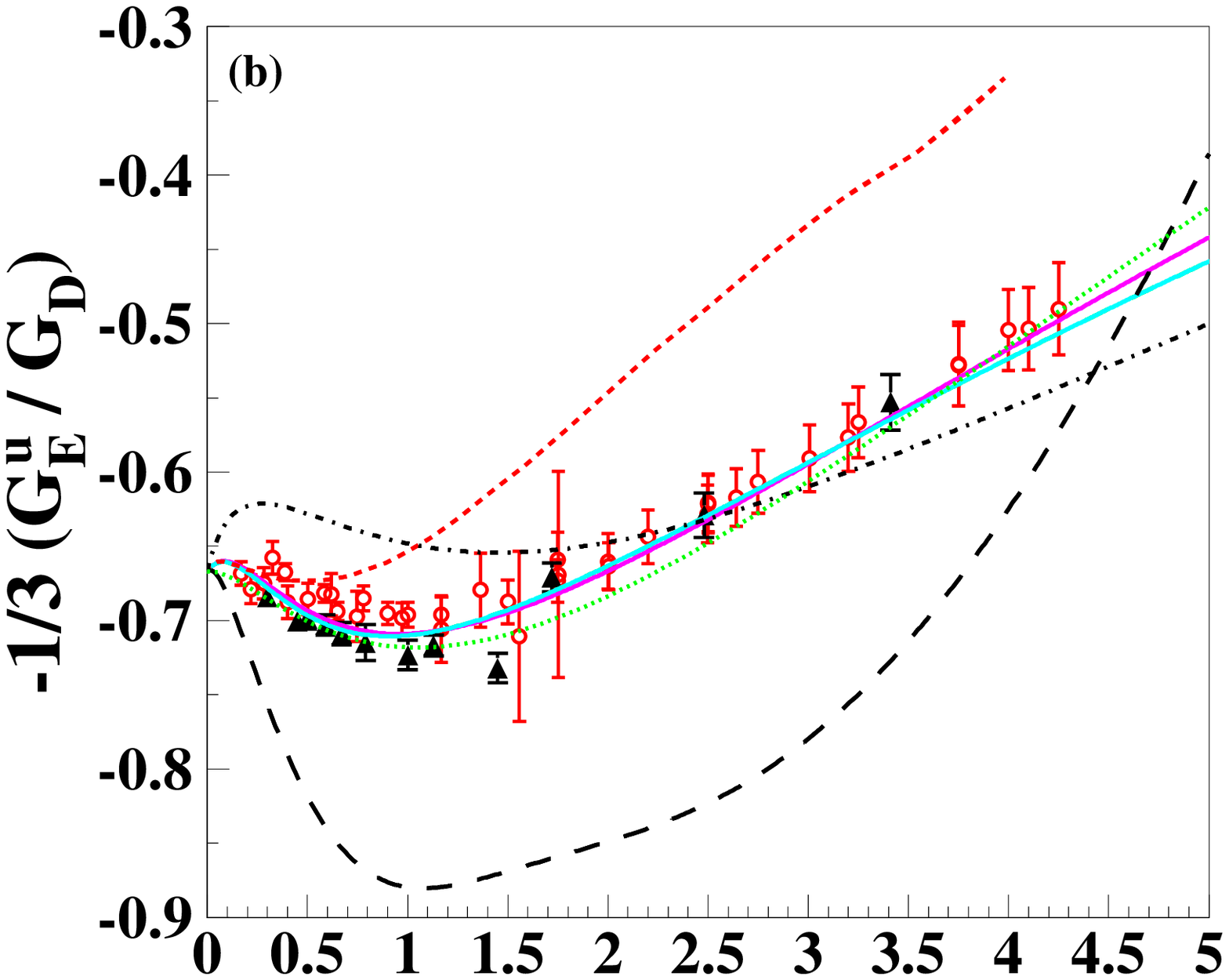}\\
\includegraphics*[width=7.9cm,height=5.5cm]{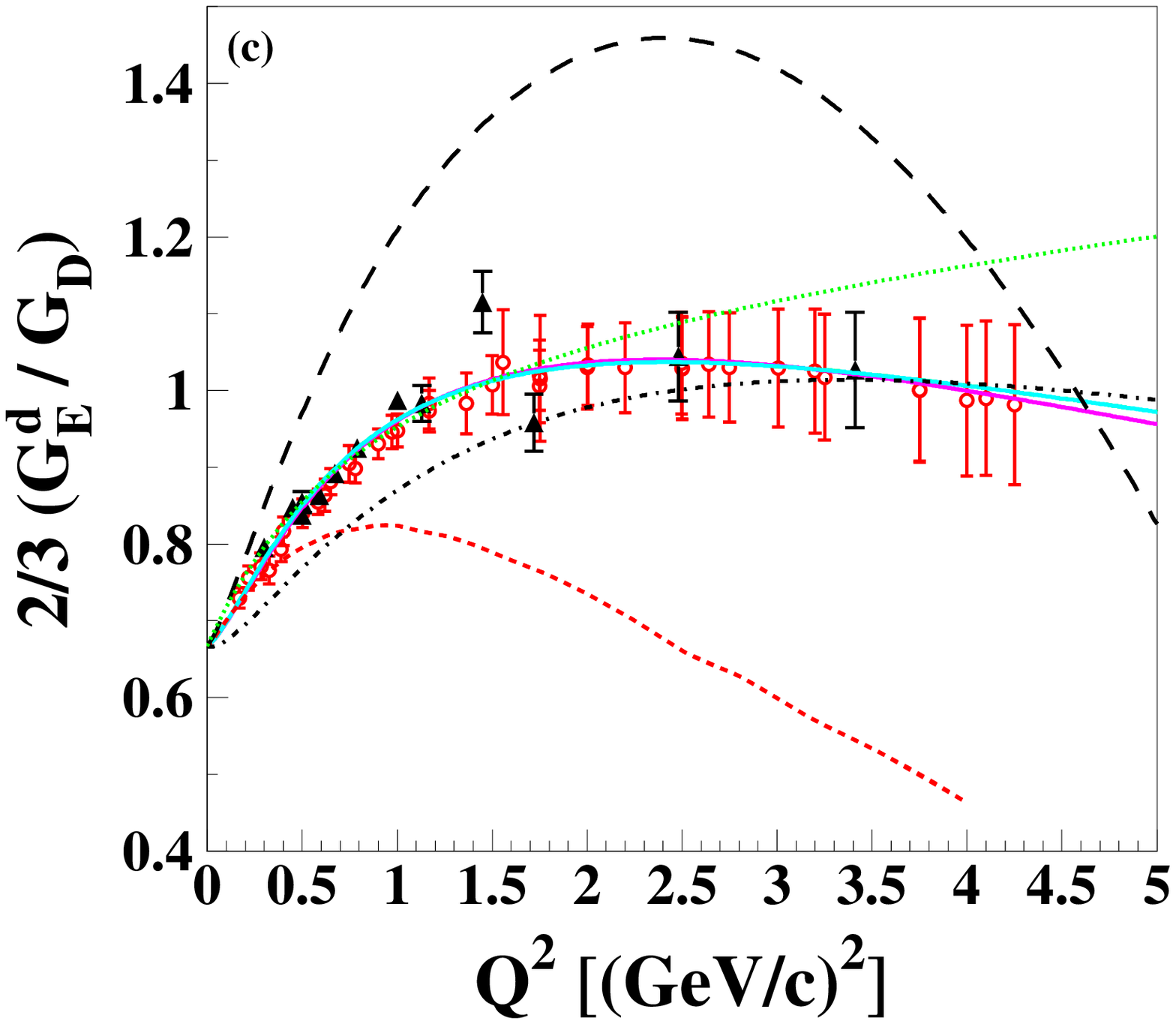}
\end{center}
\vspace{-0.5cm}
\caption{(color online) $\gen/G_D$ and its flavor-separated contributions.
Note that $G_E^u$ (middle figure) represents the up-quark contribution to the
proton, and therefore the down-quark contribution in the neutron, and so
is multiplied by the charge of the down quark. Curves and data points are the
same as in Fig.~\ref{fig:GEpFlavor}.}
\label{fig:GEnFlavor}
\end{figure}

\begin{figure}[!htbp]
\begin{center}
\includegraphics*[width=7.9cm,height=5.0cm]{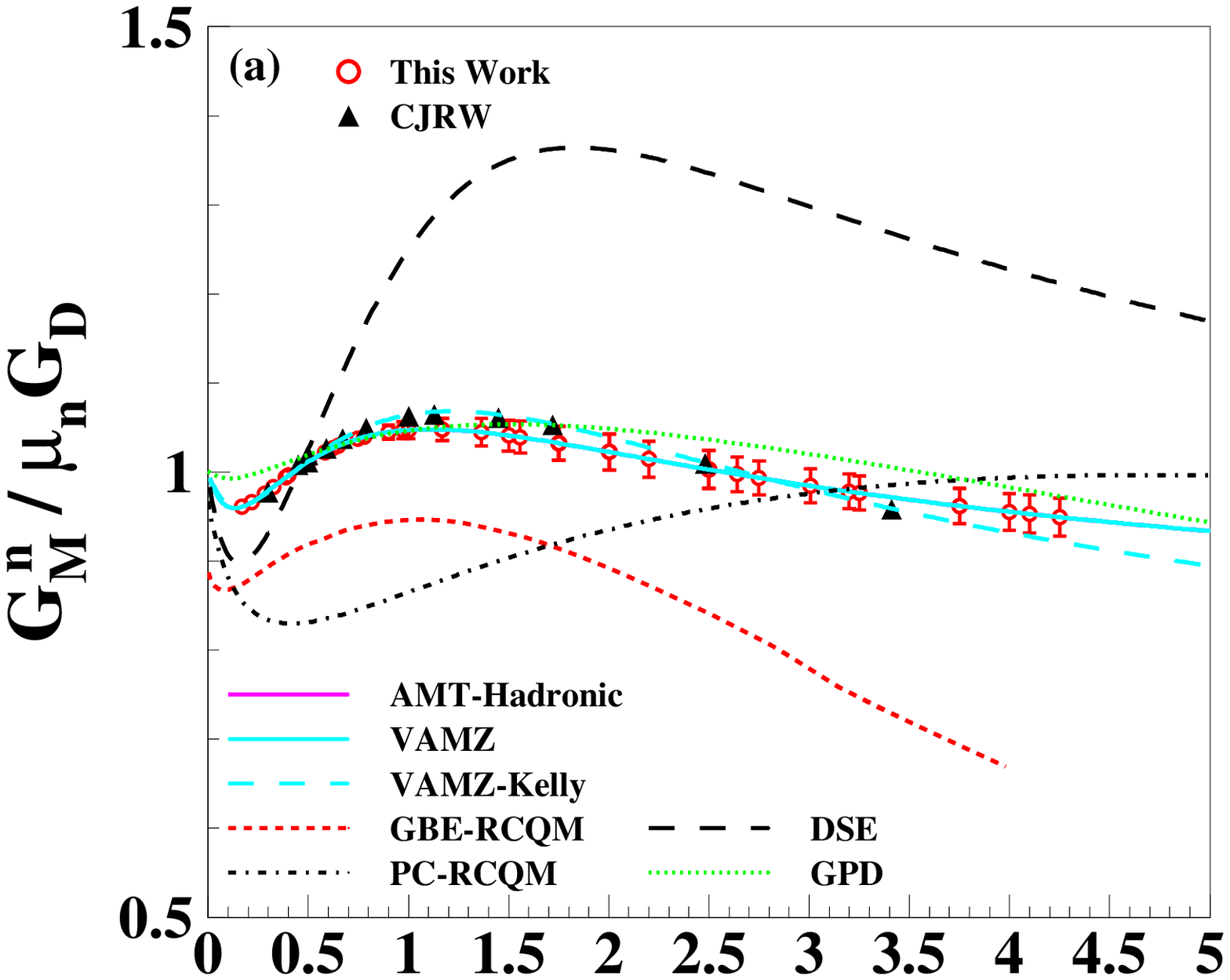}\\
\includegraphics*[width=7.9cm,height=5.0cm]{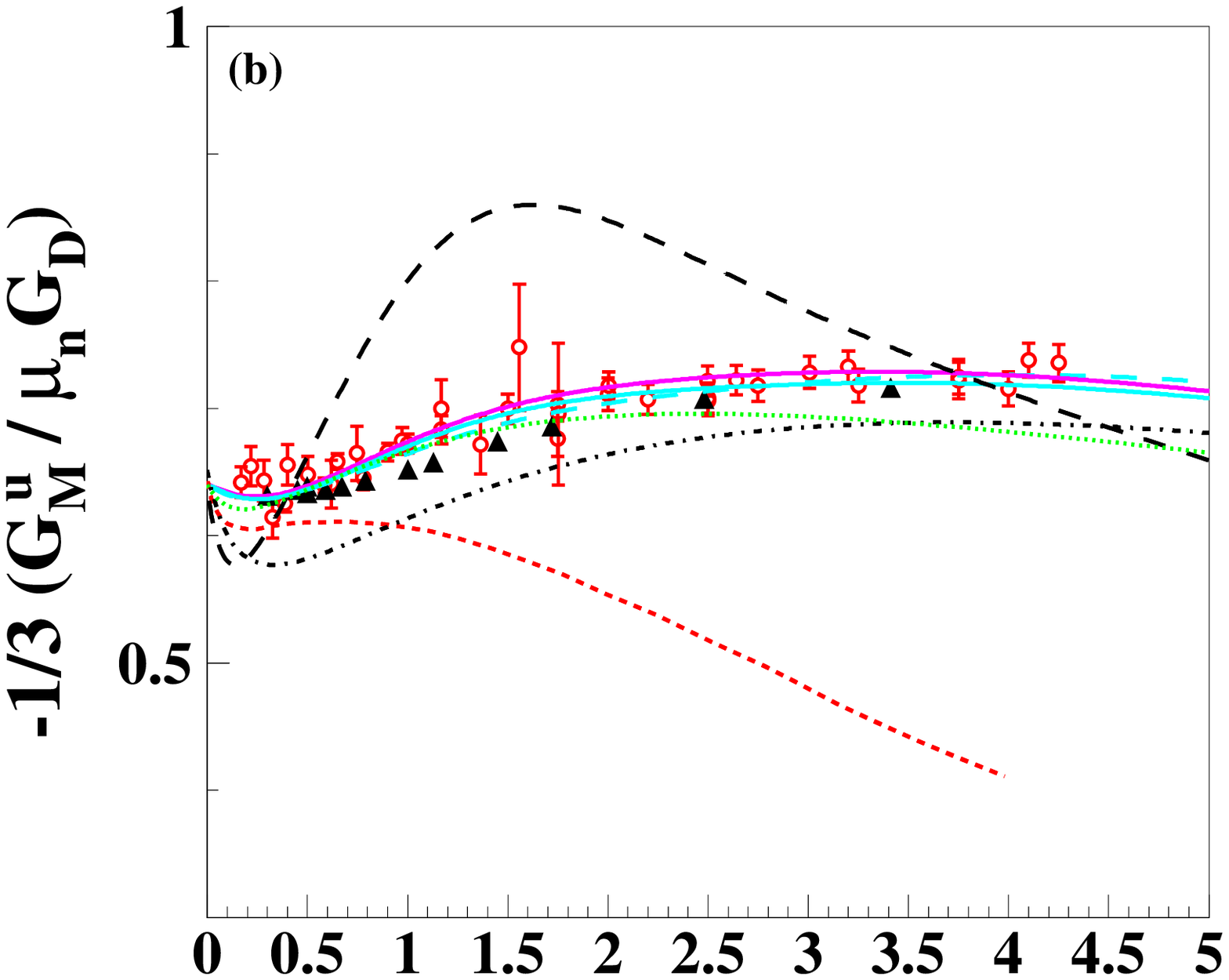}\\
\includegraphics*[width=7.9cm,height=5.5cm]{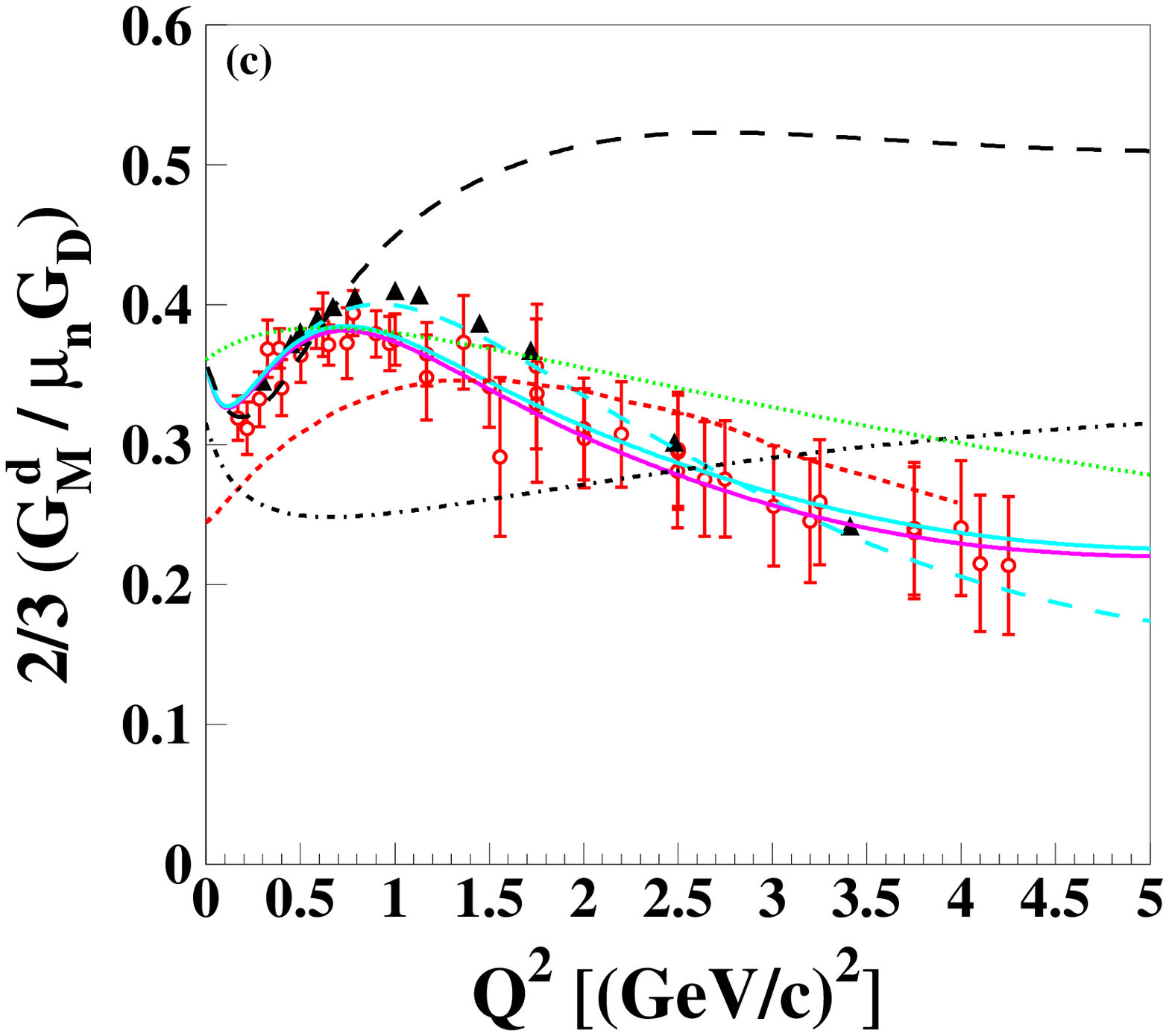}
\end{center}
\vspace{-0.5cm}
\caption{(color online) $\gmn/G_D$ and its flavor-separated contributions.
Note that $G_M^u$ (middle figure) represents the up-quark contribution to the
proton, and therefore the down-quark contribution in the neutron, and so is
multiplied by the charge of the down quark. Curves and data points are the
same as in Fig.~\ref{fig:GEpFlavor}.  In addition, we show the VAMZ-Kelly
curve which replaces our updated fit to $\gmn$ in the VAMZ curve with the
Kelly fit to show the impact of the additional $\gmn$ data included in our fit.}
\label{fig:GMnFlavor}
\end{figure}

In this section, we present the flavor-separated results for the
proton and neutron form factors.  We compare the results to the CJRW
extraction and to extractions based on the recent proton parameterizations from
Venkat~\etal~\cite{venkat11} (``VAMZ''), and from
Arrington~\etal~\cite{arrington07} (``AMT''), combined with our updated
fit to $\gmn$ and the Riordan~\etal~\cite{riordan10} parametrization of $\gen$.
Note that these are not independent extractions; they
are all based on fits to relatively up-to-date sets of form factor
measurements. The comparison to the CJRW result allows us to examine the
impact of the TPE corrections to the proton form factors, as well as the
uncertainties in the extraction of $\gep$, $\gmp$, and $\gmn$, which were
neglected in the CJRW analysis. The comparisons to the AMT and VAMZ fits
provide sensitivity to the TPE corrections at low $Q^2$, as discussed in the
previous section, as well as the impact of recent polarization transfer
data~\cite{zhan11, ron11}, which are only included in the VAMZ result. We also
examine the impact of the updated fit to $\gmn$ by showing a version of the
VAMZ extraction which uses the Kelly~\cite{kelly04} parameterization for
$\gmn$ (``VAMZ-Kelly'').  Of particular importance is the impact near
$Q^2=1$~(GeV/c)$^2$, where a tension between the CLAS data and previous
measurements yields a noticeable shift in $\gmn$, but it is not clear which
data is most correct in this region.  Finally, we compare the results to a set
of recent nucleon form factor models, described in Sec.~\ref{recent_theory}.

Figures~\ref{fig:GEpFlavor} and~\ref{fig:GMpFlavor} show the proton Sachs
form factors and their contributions from up and down quarks. Our extracted
values are included in the online supplemental
material~\cite{online_material}. The top panels show the proton form factors
used in the extraction, normalized to the dipole form, along with the
values from the CJRW analysis, the AMT and VAMZ fits, and recent calculations.

Note that for $\gep$ and $\gmp$, as well as the flavor-separated $G_M$ values,
the CJRW points have no uncertainty.  This is simply because they extract
these directly from parameterizations of $\gep$, $\gmp$, and $\gmn$ and do not
include any uncertainty in the fits. Obviously, when looking at form factors
that are independent of the $R_n$ measurements, these uncertainties cannot be
neglected.  

Our results are otherwise in relatively good agreement with the CJRW analysis. 
The different treatment of TPE corrections in the proton form factor input
yields a small difference in $\gep$ and $\gmp$ for $Q^2$ values near
0.5--1.5~(GeV/c)$^2$.  The CJRW results are in better agreement with the
improved treatment of TPE corrections in the AMT and VAMZ fits for $\gep$,
while our extraction is in better agreement for $\gmp$. At larger $Q^2$
values, both analyses and the global fits yield consistent results.

As expected, the up-quark contribution dominates both the charge and magnetic
form factor for the proton.  Examining the contributions to $\gep$, we see
that both the up- and down-quark contributions have significant deviations from
the dipole form.  At low $Q^2$ values, the increase in $G_E^u/G_D$ is
compensated by a decrease in $G_E^d/G_D$, yielding a small $Q^2$ dependence in
$\gep/G_D$.  At higher $Q^2$ values, $G_E^d$ is consistent with the dipole
form, and the decrease of $G_E^u/G_D$ leads to the overall falloff in
$\gep/G_D$. Note that the cancellation between the positive but slowly
decreasing value of $G_E^u/G_D$ and the negative but nearly constant value of
$G_E^d/G_D$ enhances the $Q^2$ dependence seen in the up-quark contribution.
Thus, the rapid linear falloff observed in $\gep/\gmp$ is a result of the
cancellation between the contributions from the up and down quarks. This
behavior is therefore connected to the difference in the up- and down-quark
distributions, rather than the overall shape of the quark distributions.

For $\gmp$, both up and down quarks have smaller deviations from the dipole
form.  At very low $Q^2$ values, both have a slightly increasing contribution,
yielding a roughly 10\% increase in $\gmp$ between $Q^2$ of 0 and 1~(GeV/c)$^2$.
Above this, the slow increase in $G_M^u$ and slow decrease in $G_M^d$ yielding
a near-perfect agreement of $\gmp$ with the dipole form up to
$Q^2$=4.5~(GeV/c)$^2$, even though both the up and down contributions have
significant deviations.

Figures~\ref{fig:GEnFlavor} and~\ref{fig:GMnFlavor} show the Sachs form factors
of the neutron along with their breakdown into up- and down-quark
contributions. In this case, our results for $\gen$ and $\gmn$ simply reflect
the values and uncertainties of the form factor parameterizations. For the
CJRW results, $\gen$ and the flavor-separated results include the
uncertainties associated with the direct measurements of $R_n$, while the
values for $\gmn$ and its up- and down-quark contributions are based entirely
on the parameterizations of the proton and neutron magnetic form factors, with
no uncertainties included.  One can see the difference between our
parameterization of $\gmn$ and the Kelly fit in Fig.~\ref{fig:GMnFlavor}(a),
where the VAMZ-Kelly fit uses Kelly fit~\cite{kelly04}, and the VAMZ result
is our updated parameterization, including the CLAS measurements.  The
difference in $\gmn$ is relatively small, but it is as large or larger than
the assumed uncertainty for $Q^2 \approx 1$~(GeV/c)$^2$ and at the largest
$Q^2$ values shown.  The impact on the up-quark contribution is negligible,
but there is a noticeable change in the extracted down-quark contribution, as
seen in Fig.~\ref{fig:GMnFlavor}(c), which is the main difference between
our extraction and the CJRW result.

Unlike in the case of the proton, the up and down quarks both yield large
contributions to the neutron form factors.  The strong $Q^2$ dependence
of $G_E^d/G_D$ at low $Q^2$, where $G_E^u/G_D$ is relatively flat, yields the
rise in $\gen$, while at larger $Q^2$ values, $G_E^d/G_D$ stops rising and
$\gen$ grows slowly compared to the dipole form due to the small $Q^2$
dependence in $G_E^u$.  As with the proton, the contributions to $\gmn$
have small deviations from the dipole, although the contribution
from the down quark, which has larger deviations, yields a small
$Q^2$ dependence in $\gmn/G_D$ at larger $Q^2$ values.

Figures~\ref{fig:GEpFlavor}--\ref{fig:GMnFlavor} also show the flavor-separated
contributions from form factor parameterizations (AMT~\cite{arrington07} and
VAMZ~\cite{venkat11}) and the calculations discussed in
Sec.~\ref{recent_theory}.  As mentioned above, the parameterizations are fits
that include much of the data included in these extractions, and so at high
$Q^2$ yield consistent results with the data.  At low $Q^2$, they help show
the impact of two-photon exchange corrections which are neglected in the
CJRW extraction and treated in a way that is less reliable at low $Q^2$
in our analysis.

The calculations all give a reasonable qualitative description of the 
up- and down-quark contributions, showing $G_E^u/G_D$ rising at high
$Q^2$, $G_E^d/G_D$ rising and then leveling off or falling, and relatively
little $Q^2$ dependence in the up- and down-contributions to $G_M/G_D$.
The GPD model~\cite{hernandez2012} gives the best description of the data,
with only small deviations at large $Q^2$.  This is not surprising as it
includes an essentially complete set of data in fitting the GPDs, and uses
a GPD parameterization with sufficient flexibility to reproduce the data. 
The PC-RCQM result~\cite{cloet2012} also does a good job in reproducing the
behavior of the data, although with significantly larger deviations in $G_M^d$
(and $G_M^n$) than in the other form factors.  The GBE-RCQM
calculation~\cite{rohrmoser11} does a fairly good job of reproducing the data
at small low $Q^2$ values, but above 1~(GeV/c)$^2$, shows large deviations
from the data in both the flavor-separated and the proton and neutron form
factors.  While it does not reproduce the data as well as the GPD or PC-RCQM
curves, it is a parameter-free calculation, making the overall agreement
rather remarkable.

The DSE calculation~\cite{cloet09} has significant deviations at both low
and high $Q^2$.  However, in the DSE approach, the nucleon mass and quark
magnetic moments are not forced to reproduce the physical values, as is
expected because additional contributions from pseudoscalar mesons, the ``pion
cloud'' contributions, excluded in this calculation, will bring these closer
to the physical values.  The missing pseudoscalar meson contributions are
expected to modify the behavior at lower $Q^2$ values, while the unphysical
nucleon mass in this model may modify the comparison at higher $Q^2$ values.
For this comparison, we used the physical nucleon masses and magnetic moments
to partially account for this difference.  However, while taking the physical
magnetic moments yields the correct limit as $Q^2 \to 0$, it may worsen the
agreement at larger $Q^2$ values, where the pion cloud contributions are not
expected to be as important.  Note that the DSE calculation has most of its
parameters fixed based on the calculation of light mesons, and the only
additional parameters for the nucleon calculation are the diquark radii, taken
to be commensurate with the pion's charge radius.  So as with the GBE-RCQM
calculation, the result is not adjusted to improve agreement with the form
factor data.

\begin{figure}[!htbp]
\begin{center}
\includegraphics*[width=7.9cm,height=5.5cm]{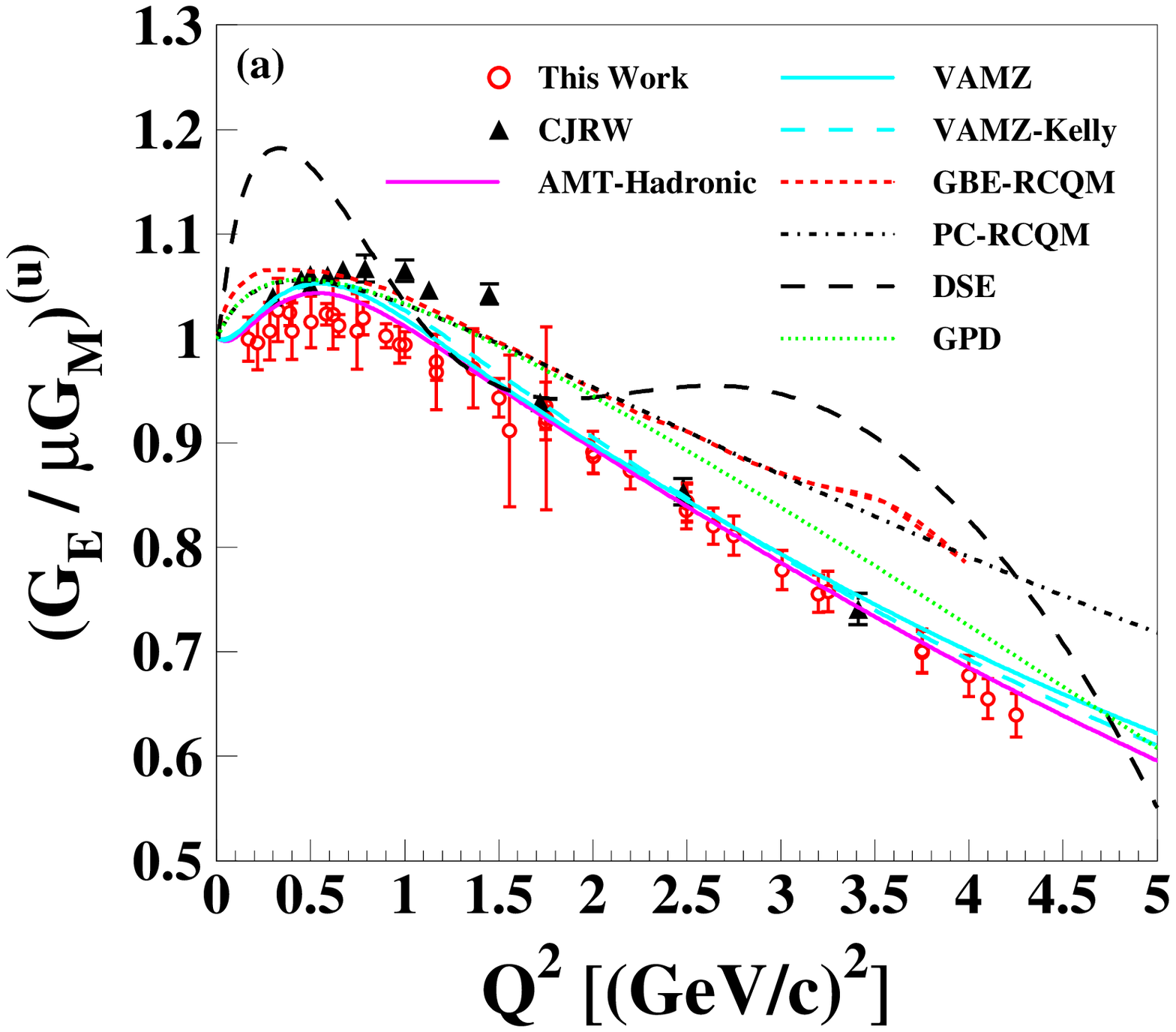}\\
\includegraphics*[width=7.9cm,height=5.5cm]{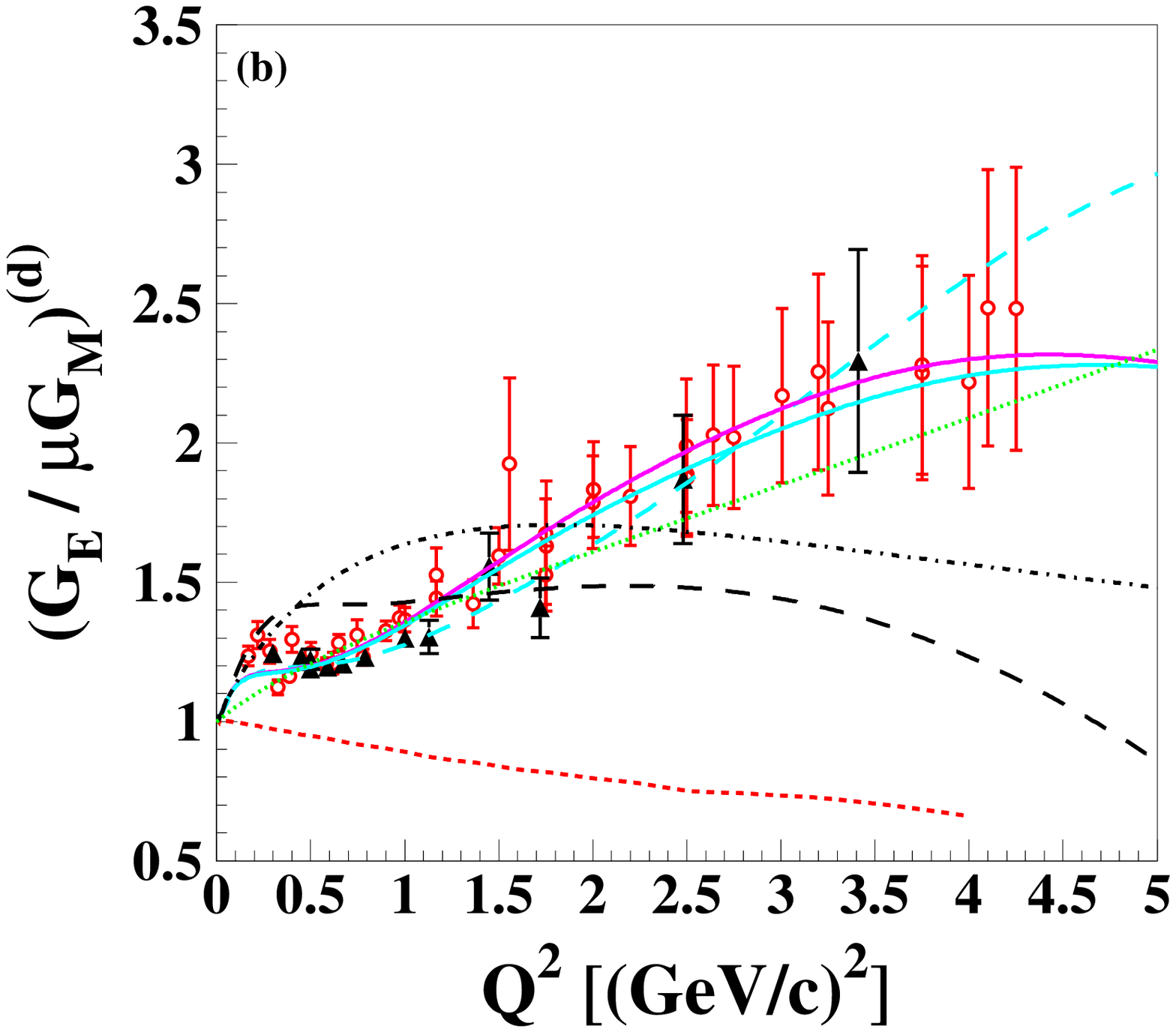}
\end{center}
\vspace{-0.5cm}
\caption{(color online) The ratio $G_E/G_M$ for the up- and down-quark
contributions, normalized to unity at $Q^2=0$.}
\label{fig:GeGm}
\end{figure}

Figure~\ref{fig:GeGm} shows the ratio $G_E/\mu G_M$ for both the up and down
quarks.  As before, the extractions for the up-quark contribution disagree
somewhat at low $Q^2$, with the fits including a more complete TPE treatment
agreeing better with CJRW below 1~(GeV/c)$^2$, and our results above.  For the
down-quark contributions, the primary disagreement comes from the difference
in the $\gmn$ results, whose impact can be seen by comparing the VAMZ and
VAMZ-Kelly curves.

For the up quark, $G_E/G_M$ has a roughly linear falloff at large $Q^2$, but
the decrease is slower than seen for the proton.  For the down quark, a
completely different behavior is seen, with $G_E/G_M$ increasing in magnitude
with $Q^2$. This behavior is not present in any of the calculations except for
the GPD model, which has a sufficiently flexible parameterization of the GPDs
to reproduce all of the form factors.

We now examine the Dirac and Pauli form factors, with an emphasis on the
high-$Q^2$ behavior of the flavor-separated contributions.  While
perturbative-QCD behavior should set in at large enough $Q^2$ values, the data
do not extend into the region where one expects these asymptotic predictions
to be valid.  Nonetheless, it has been observed that approximate scaling
of the form factors often sets in at lower $Q^2$ values.  By taking out the
predicted high-$Q^2$ behavior, we can more easily see differences in
the $Q^2$ dependences of the various contributions to the form factors.

\begin{figure}[!htbp]
\begin{center}
\includegraphics*[width=7.9cm,height=5.0cm]{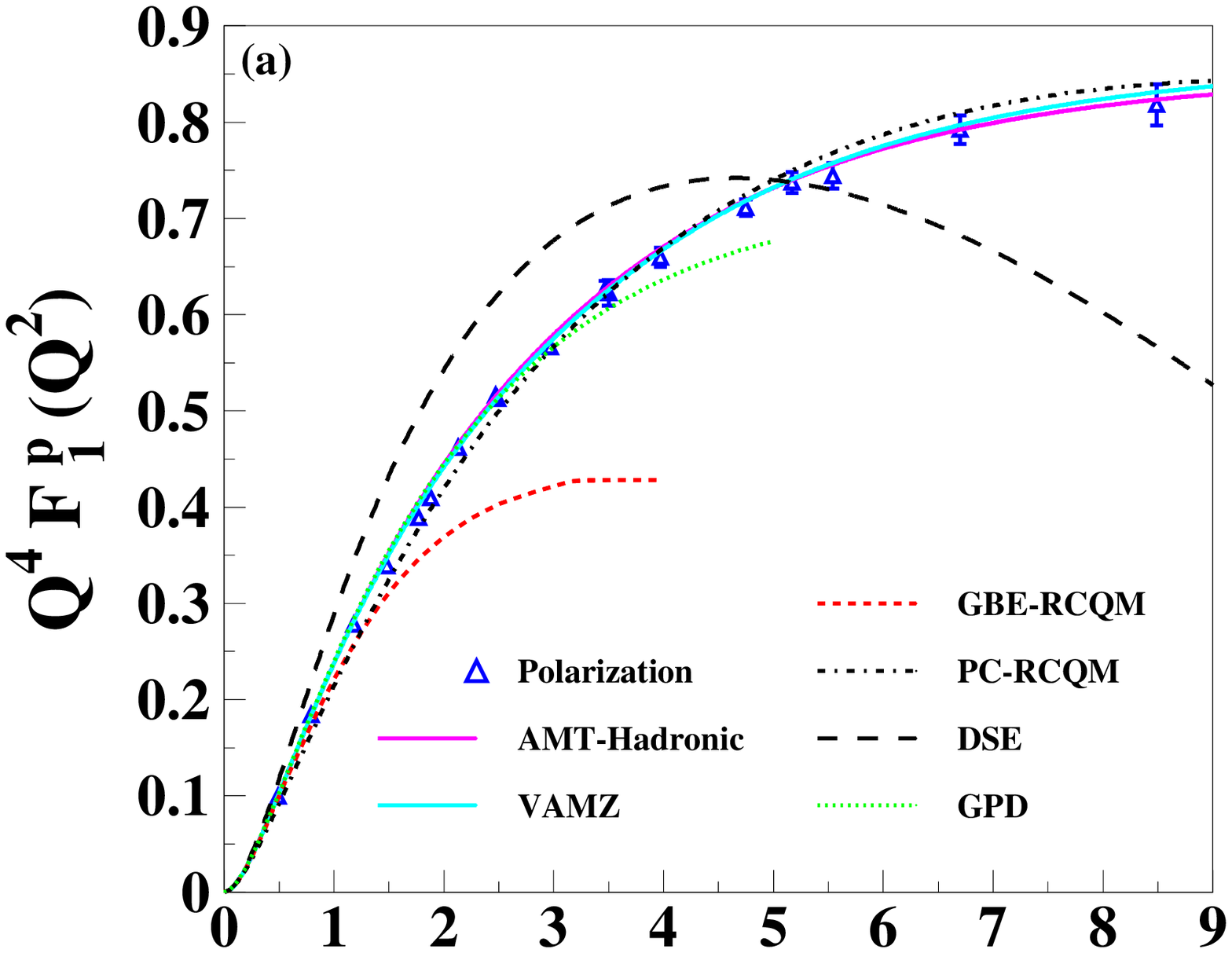}\\
\includegraphics*[width=7.9cm,height=5.0cm]{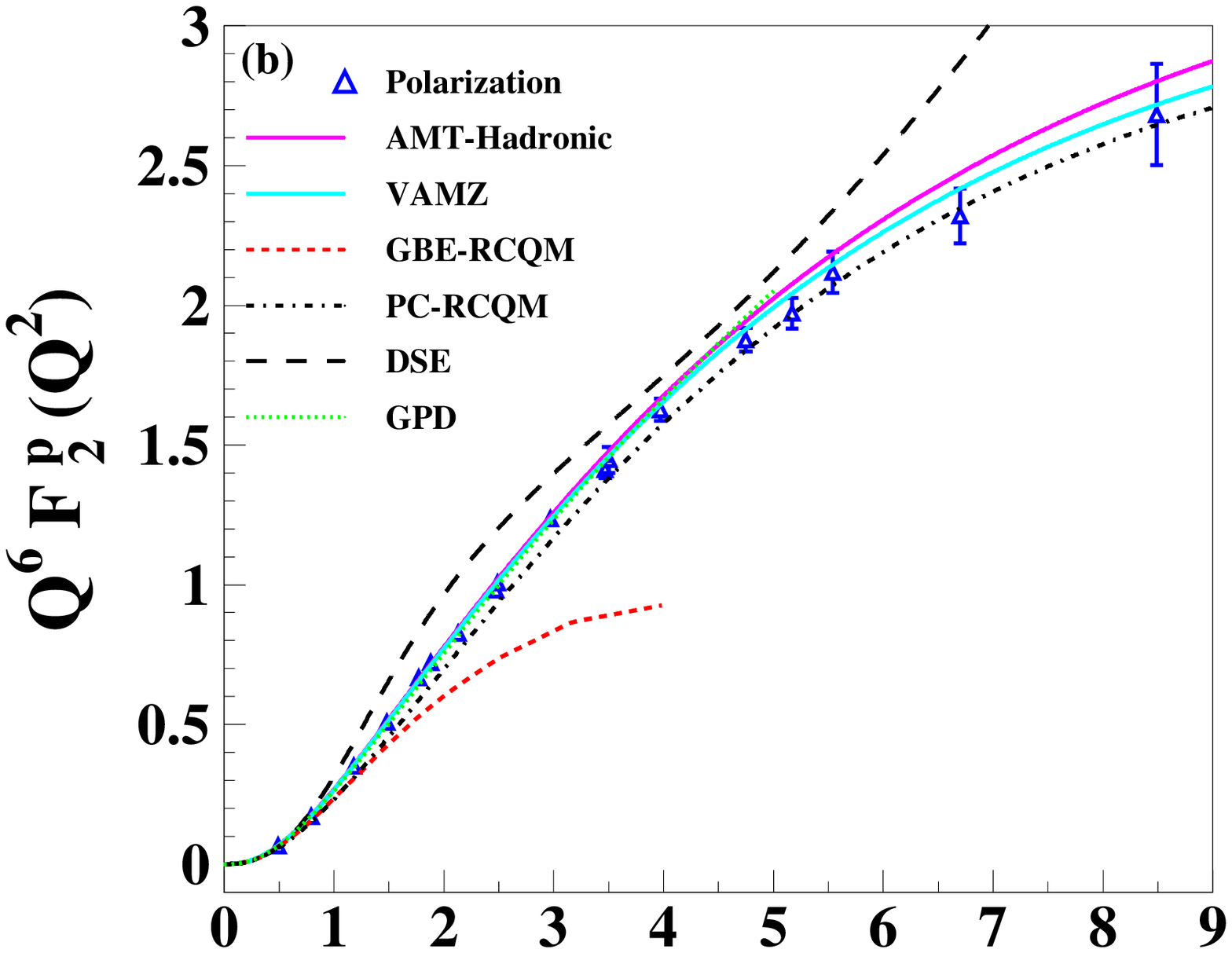}
\includegraphics*[width=7.9cm,height=5.5cm]{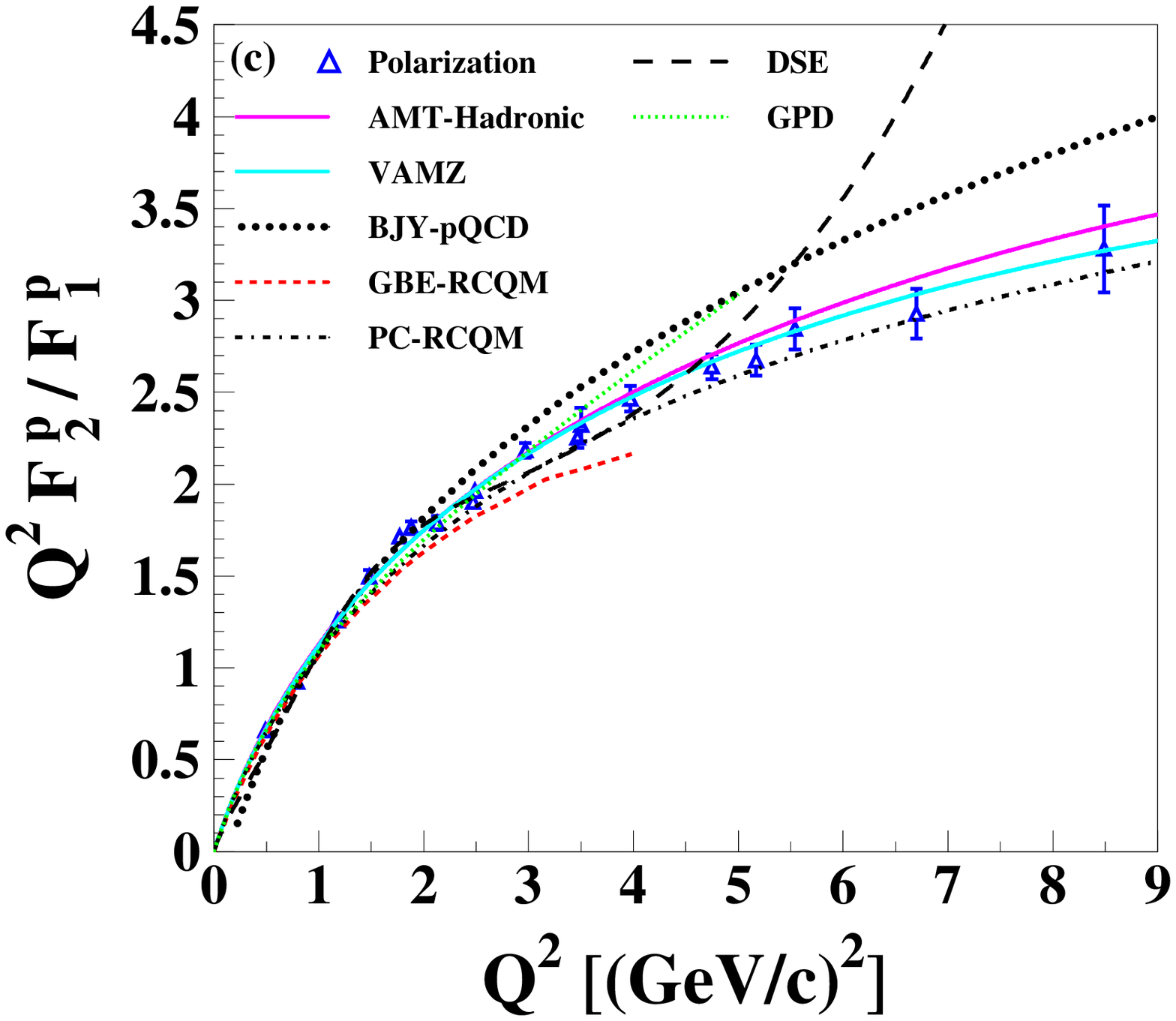}
\end{center}
\vspace{-0.5cm}
\caption{(color online) $Q^4 F^p_1$ [top], $Q^6 F^p_2$ [middle], and $Q^2
F_2^p/F_1^p$ [bottom] from polarization measurements of $R_p$~\cite{punjabi05,
puckett10, puckett12} and $\gmp$ parameterization of Kelly~\cite{kelly04}. 
Also shown are the AMT~\cite{arrington07} and VAMZ~\cite{venkat11} fits, and
the calculations discussed in Sec.~\ref{recent_theory}.  We also show the
modified pQCD scaling fit from Ref.~\cite{belitsky03} with $\Lambda$=300~MeV,
labeled ``BJY-pQCD''.}
\label{fig:F1pF2p}
\end{figure}

Figure~\ref{fig:F1pF2p} shows the proton Dirac and Pauli form factors and
their ratio.  In all cases, the leading perturbative (``pQCD'') $Q^2$
dependence~\cite{lepage79} is removed by scaling the results by powers of
$Q^2$. While pQCD suggests scaling behavior of the form $F^p_1 \propto
Q^{-4}$, $F^p_2 \propto Q^{-6}$, and $F^p_2/F^p_1 \propto Q^{-2}$, the data
clearly do not support such scaling as both $Q^4F^p_1$ and $Q^6F^p_2$ 
increase with $Q^2$.  While Rosenbluth extractions which did not include TPE
corrections observed scaling behavior in the flattening of the ratio $Q^2
F_2^p/F_1^p$, the high-$Q^2$ recoil polarization measurements~\cite{gayou02,
puckett10} show that the pQCD scaling behavior is not observed. Note that
while both $F_1^p$ and $F_2^p$ deviate from the scaling predictions, these
deviations are different enough that the ratio also deviates from the pQCD
expectation.  For the ratio $F^p_2/F^p_1$, we also show a curve based on
updated result of Ref.~\cite{belitsky03} which includes an additional 
logarithmic term that goes like $\ln(Q^2/\Lambda^2)$.  The data can be well
reproduced with a value of $\Lambda$ near 300~MeV.  However, it is not clear
that the data in this region should be described by perturbative behavior,
even with logarithmic corrections, and the best fit value of $\Lambda \approx
300$~MeV appears to be too small to be an appropriate value for
$\Lambda$~\cite{arrington07a}.

\begin{figure}[!htbp]
\begin{center}
\includegraphics*[width=7.9cm,height=5.0cm]{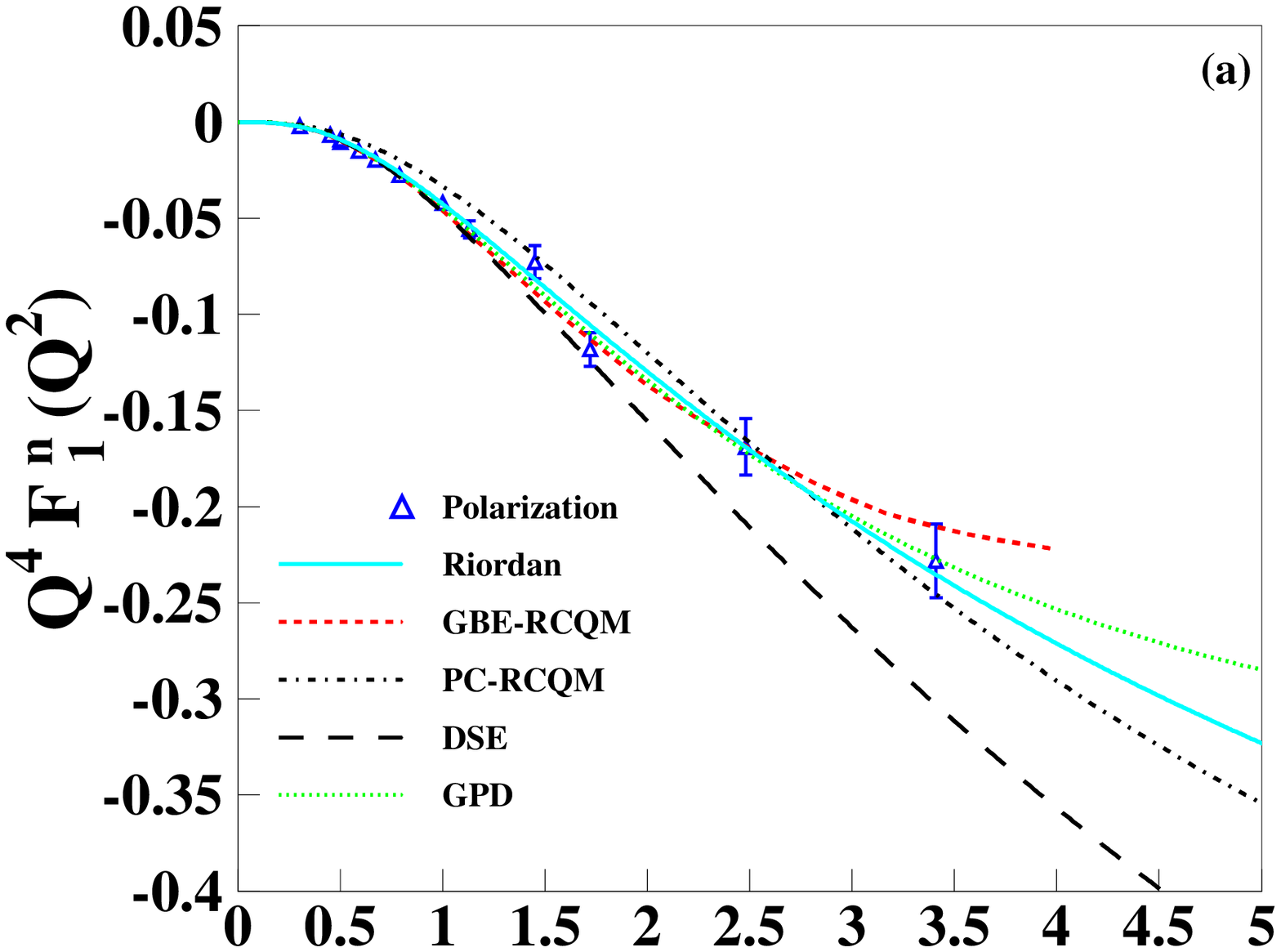}\\
\includegraphics*[width=7.9cm,height=5.0cm]{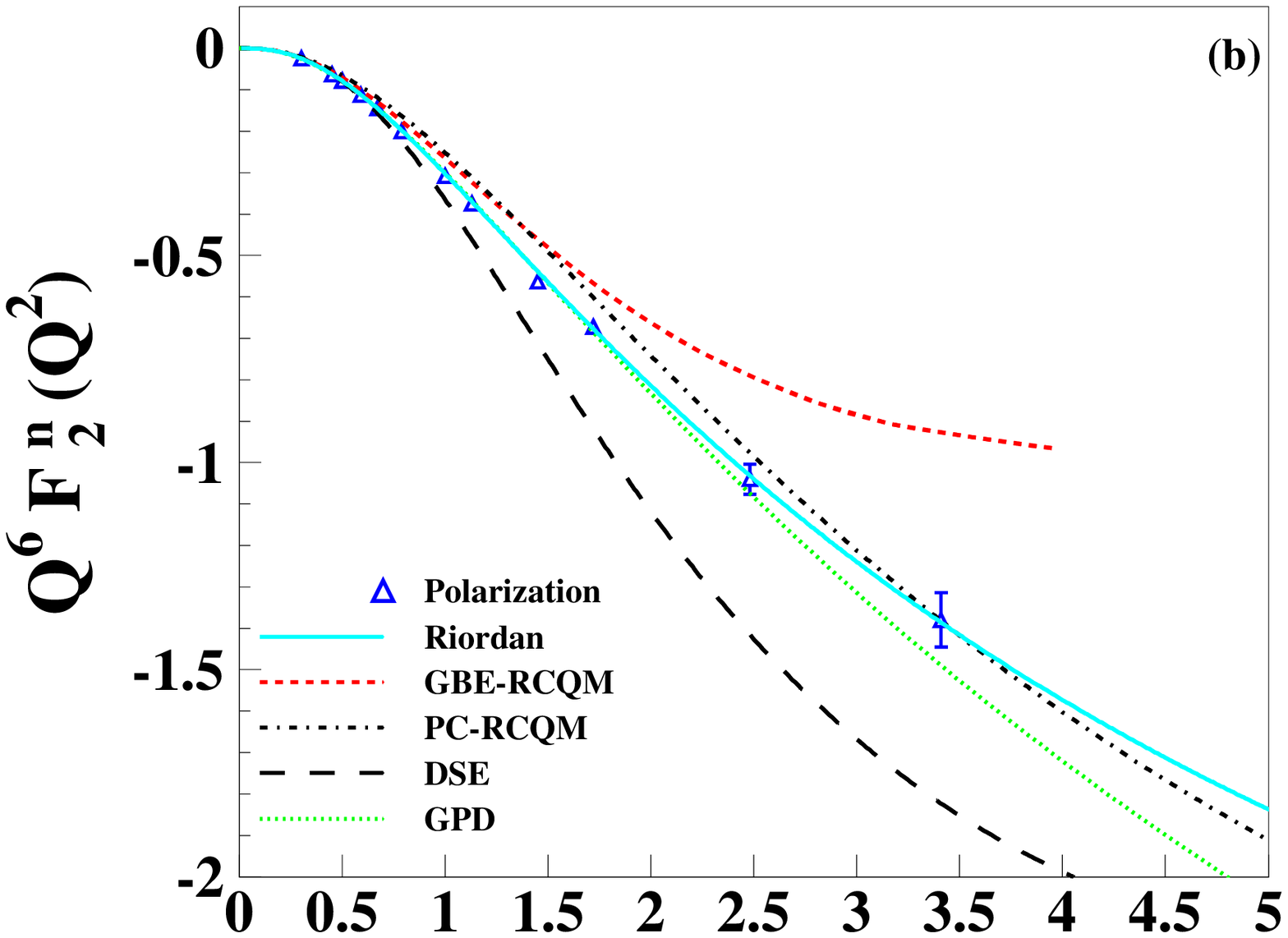}\\
\includegraphics*[width=7.9cm,height=5.5cm]{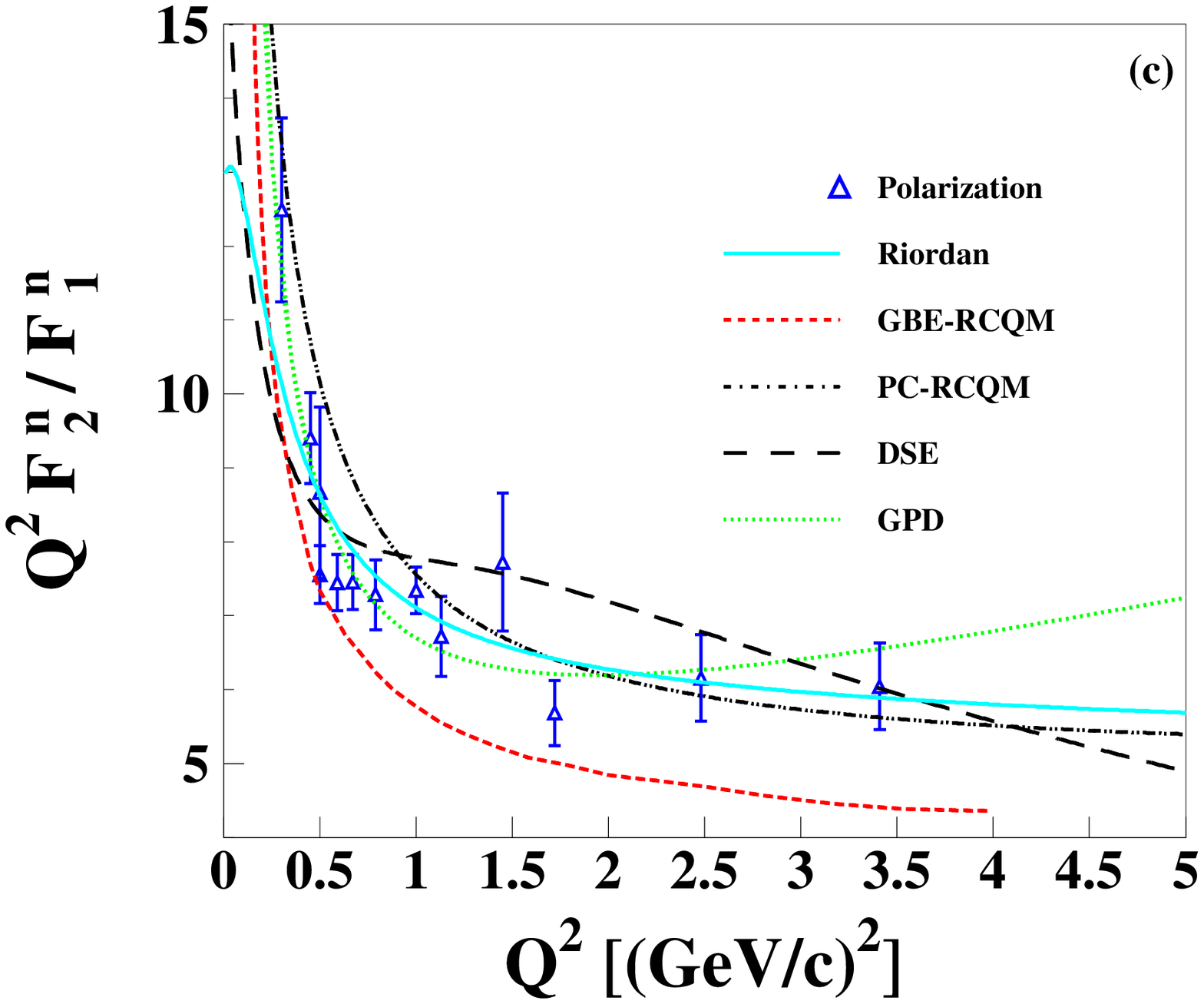}
\end{center}
\vspace{-0.5cm}
\caption{(color online) $Q^4 F^n_1$ [top], $Q^6 F^n_2$ [middle], and
$Q^2 F_2^n/F_1^n$ [bottom] from polarization measurements of
$R_n$~\cite{riordan10, zhu01, bermuth03, warren04, glazier05, plaster06}
and the $\gmn$ parameterization of Kelly~\cite{kelly04}.  The curve 
labeled ``Riordan'' uses the Riordan, et al. fit to $R_n$~\cite{riordan10}
combined with our
updated fit to $\gmn$, as described in Sec.~\ref{formfactor_tpe}.  Also shown
are the calculations presented in Sec.~\ref{recent_theory}.}
\label{fig:F1nF2n}
\end{figure}

\begin{figure}[!htbp]
\begin{center}
\includegraphics*[width=7.9cm,height=5.0cm]{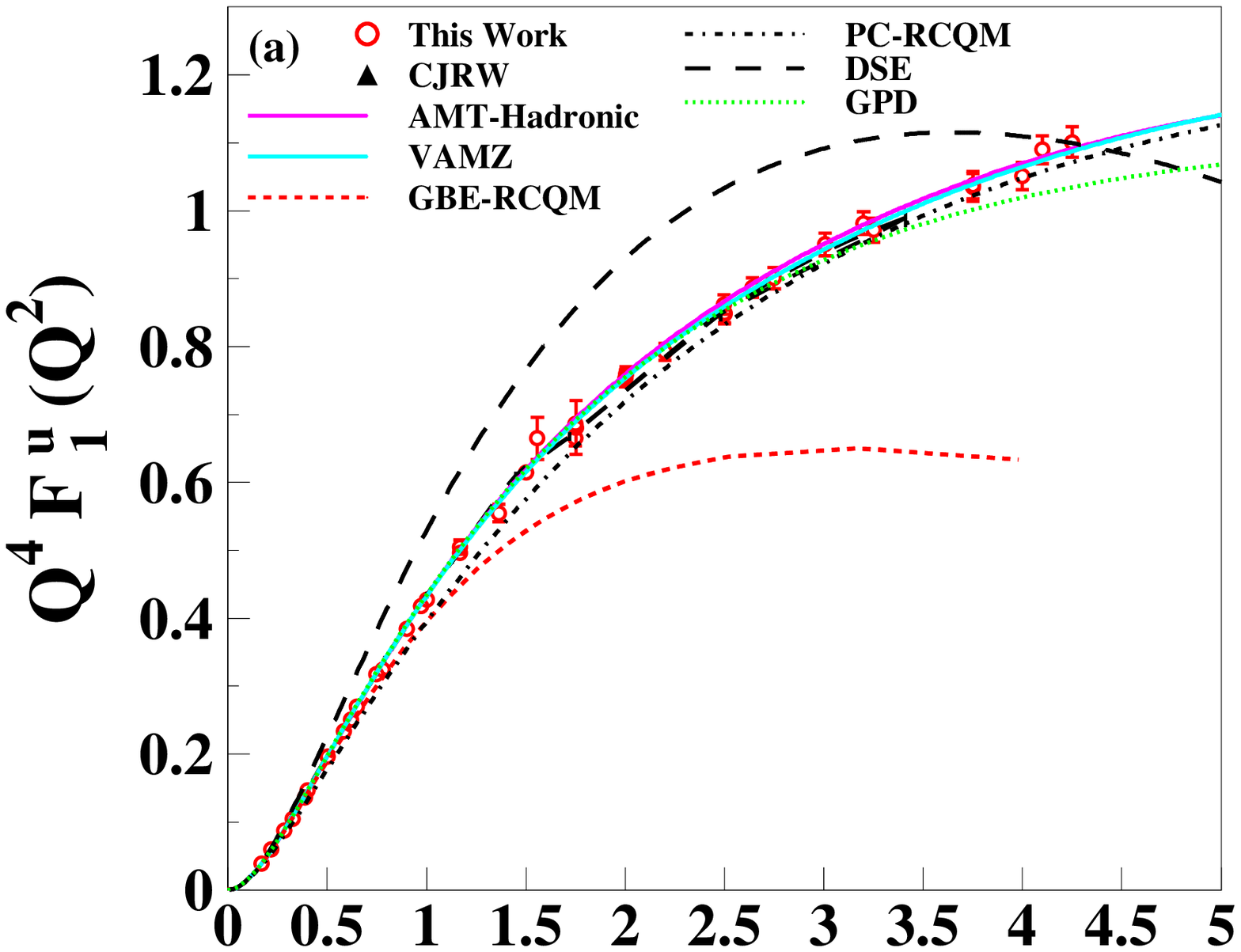}\\
\includegraphics*[width=7.9cm,height=5.0cm]{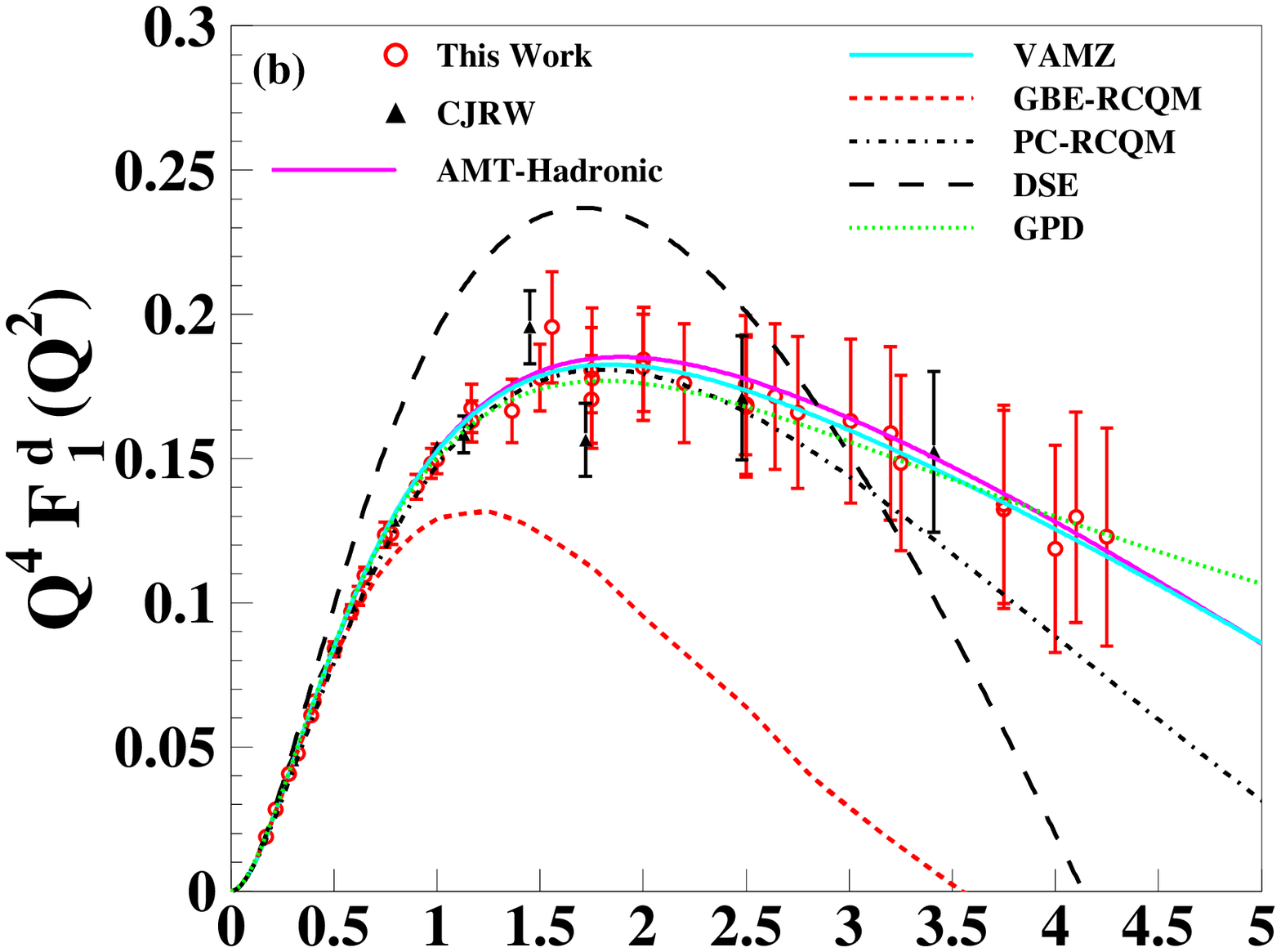}\\
\includegraphics*[width=7.9cm,height=5.5cm]{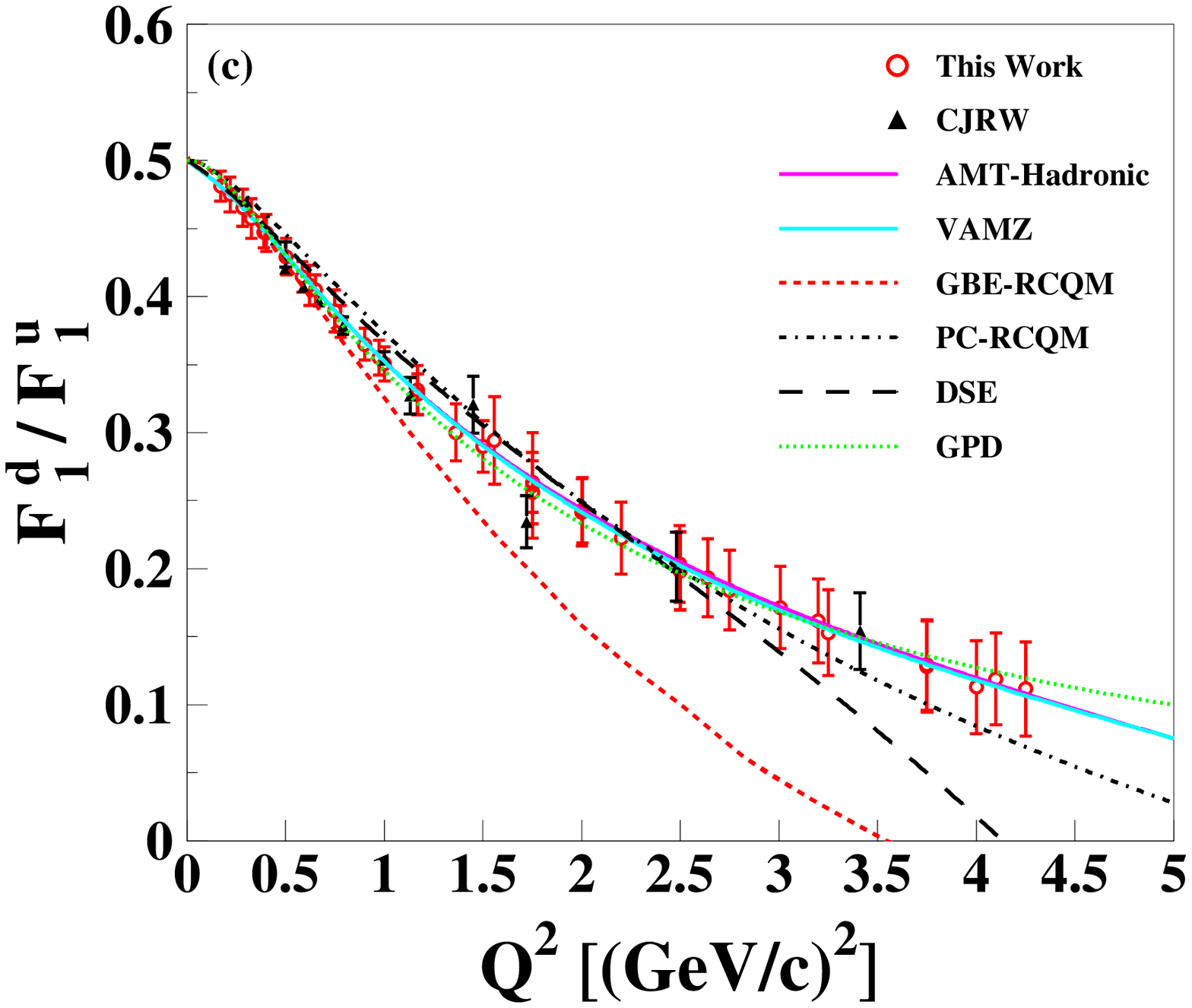}
\end{center}
\vspace{-0.5cm}
\caption{(color online) The up-quark [top] and down-quark [middle] contribution
to the Dirac form factor multiplied by $Q^4$, along with their ratio [bottom].}
\label{fig:Q4F1uF1d}
\end{figure}

\begin{figure}[!htbp]
\begin{center}
\includegraphics*[width=7.9cm,height=5.0cm]{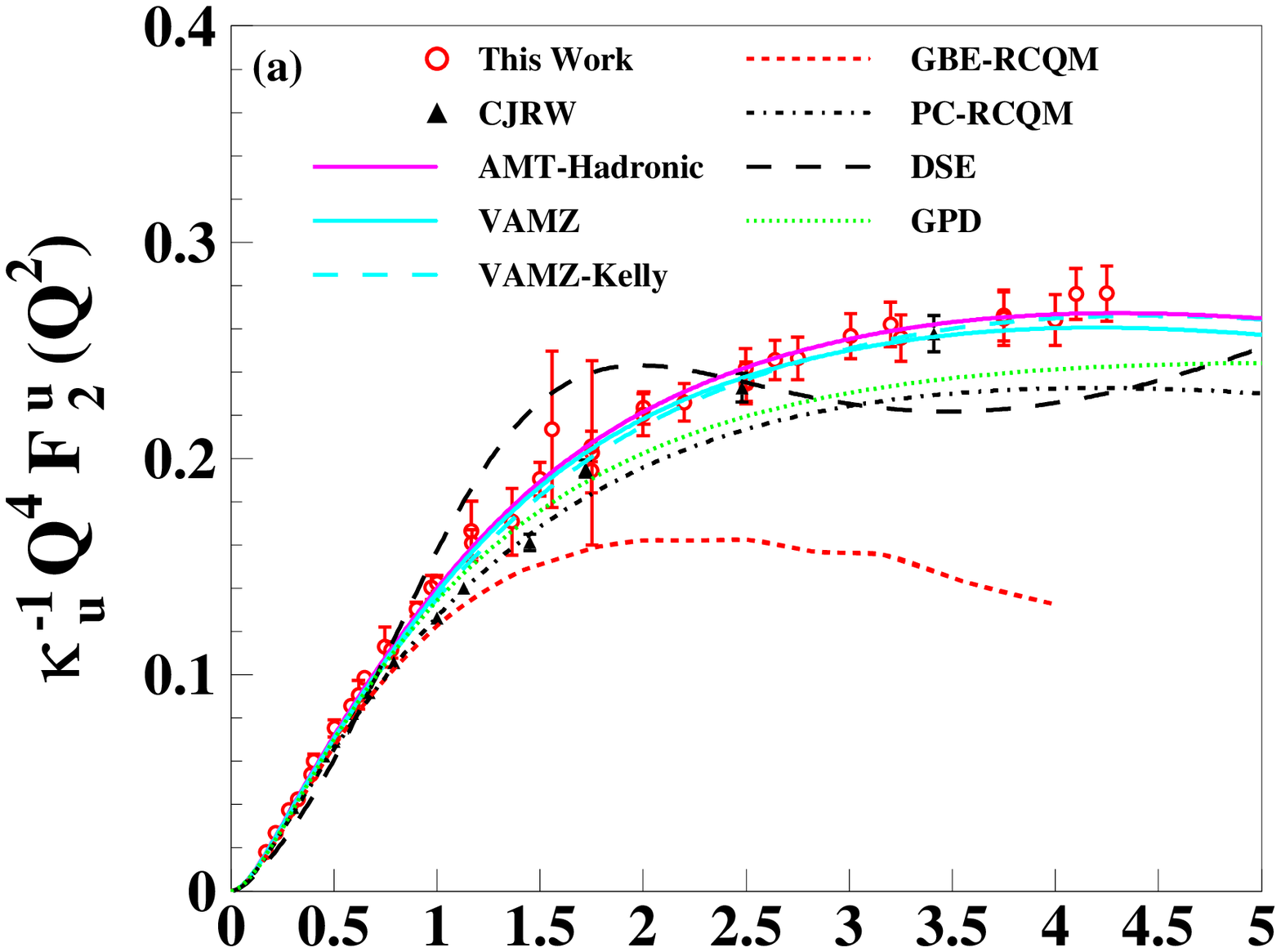}\\
\includegraphics*[width=7.9cm,height=5.0cm]{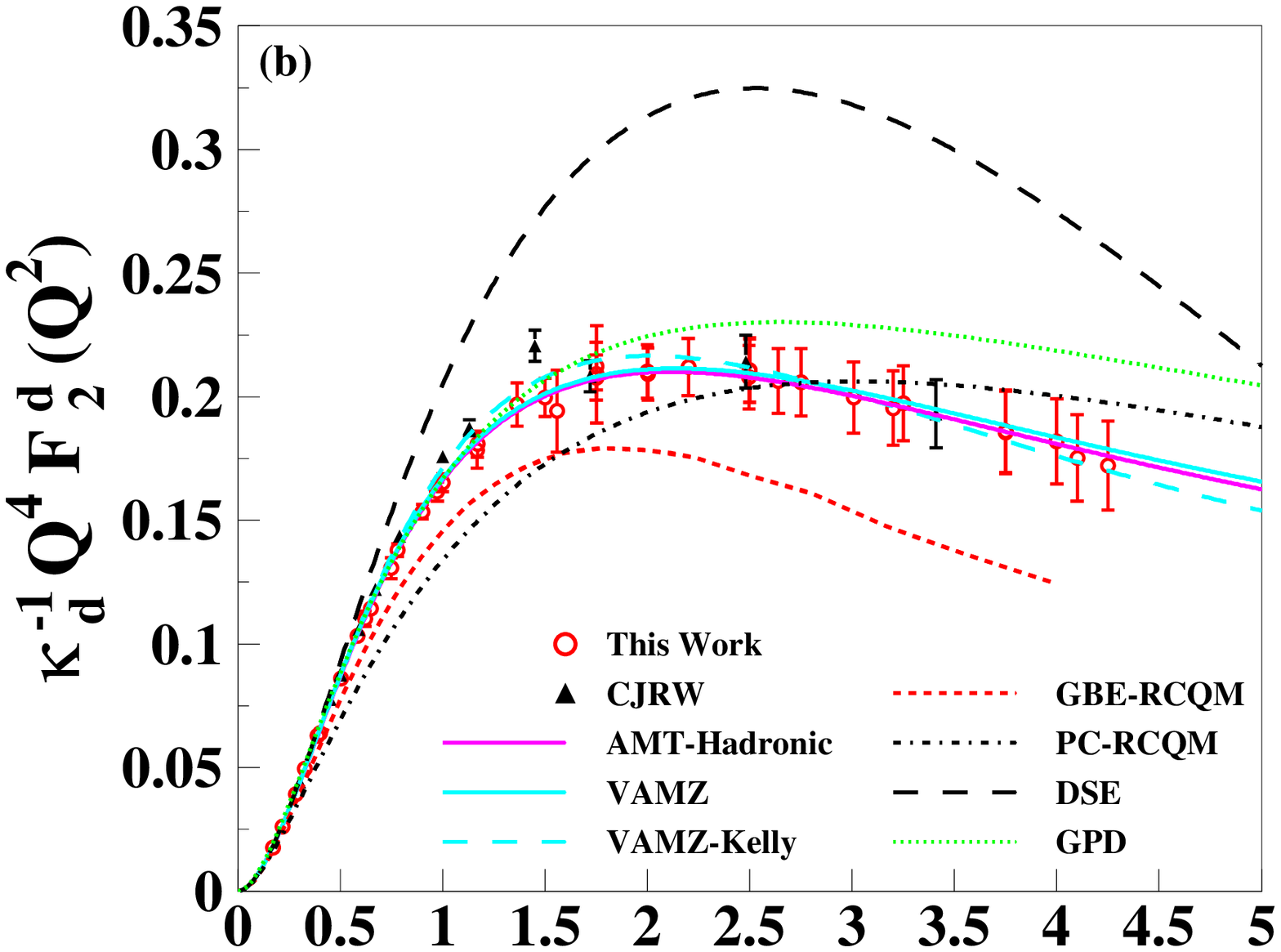}\\
\includegraphics*[width=7.9cm,height=5.5cm]{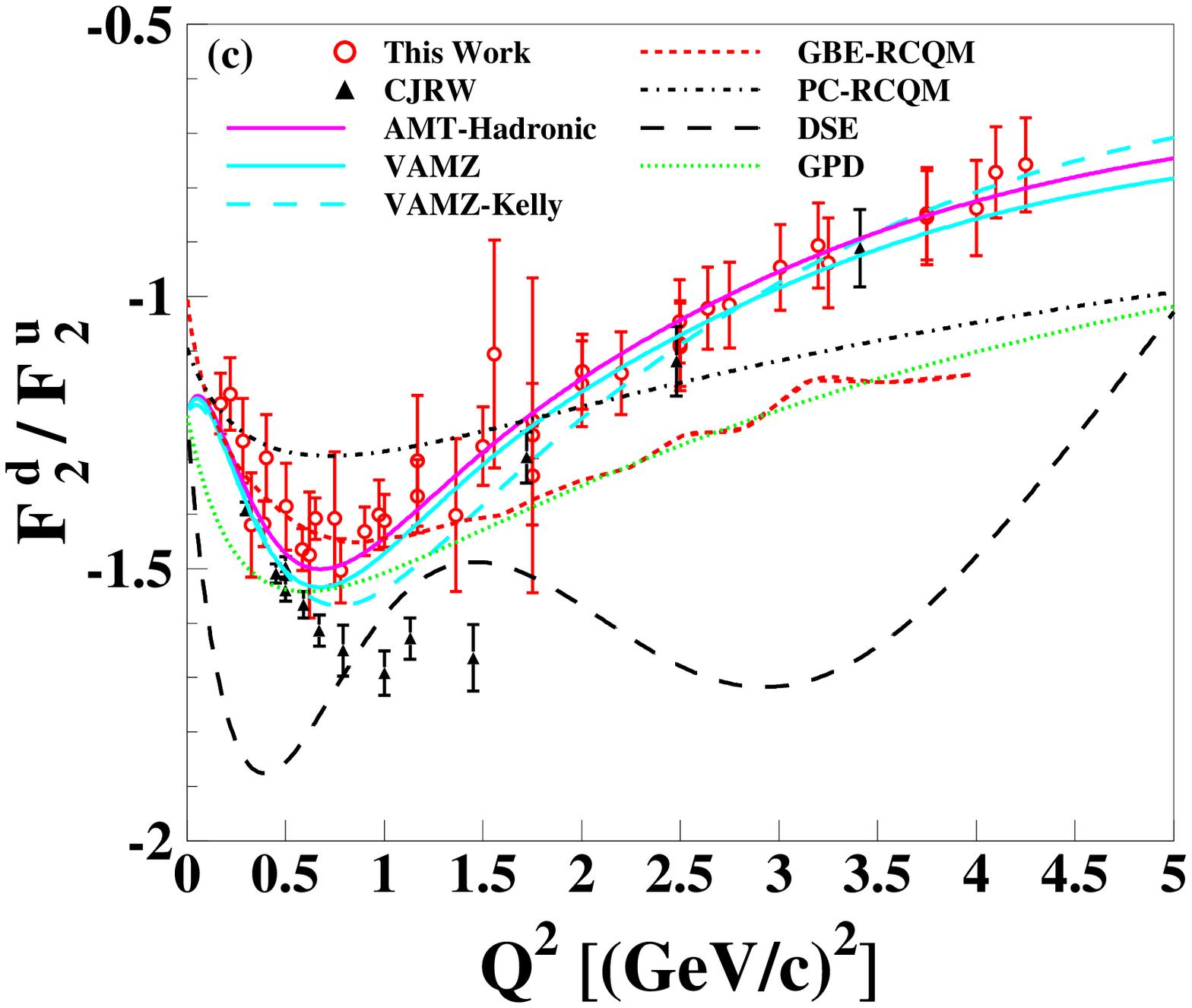}
\end{center}
\vspace{-0.5cm}
\caption{(color online) The up-quark [top] and down-quark [middle] contribution
to the Pauli form factor, multiplied by $\kappa_{(u,d)}^{-1} Q^4$, along with
the ratio $F_2^d/F_2^u$ [bottom].}
\label{fig:Q4F2uF2d}
\end{figure}

Figure~\ref{fig:F1nF2n} shows the same quantities for the neutron. Again,
neither $F_1^n$ nor $F_2^n$ are consistent with the pQCD predictions.  However,
unlike the proton, the ratio $Q^2 F_2^n/F_1^n$ is consistent with a constant value
above $Q^2$=1.5~(GeV/c)$^2$, as noted in the CJRW analysis~\cite{CJRW2011},
although the precision of the data do not set tight constraints on the $Q^2$
dependence.

The DSE and GBE-RCQM calculations generally show similar deviations from
experiment for the Dirac and Pauli form factors (for both the proton and the
neutron), yielding better agreement with the ratio than the individual
form factors.  This also helps explain why these calculations are in somewhat
better agreement with $G_E$ than $G_M$, as $G_E$ relates to the difference
between $F_1$ and $F_2$, yielding a partial cancellation of the 
deviations from the data.  At large $Q^2$ values, the deviations in the DSE
calculation grow and are of the opposite sign for $F_1$ and $F_2$. However,
the calculation does not adjust the diquark radius to better reproduce the
proton form factor, and the deviations from the data can be significantly
improved by increasing the diquark radius~\cite{cloet09}.  At large $Q^2$
values, the behavior of the ratio is also extremely sensitive to the
dressed-quark mass function, as discussed in Sec. III of Ref.~\cite{wilson12},
due to the significant cancellation between $F_1$ and $F_2$ contributions.

We now turn to the flavor-separated contributions of $F_1$ and $F_2$. The
extracted form factors are shown in Figs.~\ref{fig:Q4F1uF1d}
and~\ref{fig:Q4F2uF2d}, and the values are provided in the supplemental online
material~\cite{online_material}. Our values for $F^u_2$ are somewhat higher
than the CJRW extractions at low $Q^2$, and the impact of this difference is
seen clearly in the $F^d_2/F^u_2$ ratio.  In this case, there is a small
contribution associated with taking the updated $\gmn$ parameterization, but
the larger effect comes from the impact of TPE corrections on the proton form
factors.  While the $\gen$ uncertainties have the largest impact, the
additional contribution from the proton and $\gmn$ yield a non-negligible
increase in the total uncertainties.

In the CJRW analysis~\cite{CJRW2011} it was reported that both $F^d_1$ and
$F^d_2$ strikingly exhibit $1/Q^4$ scaling above $Q^2$=1.0~(GeV/c)$^2$, in
contrast to the up-quark form factors which continued to rise relative to the
down-quark values.  This was in agreement with the predictions for the moments
of the generalized parton distributions reported in Ref.~\cite{diehl05},
although these again are based on fits to data sets which include the nucleon
form factors, except for the most recent $\gen$ data.  Both our results and
the global parameterizations suggest that the down-quark contributions are
falling slightly with respect to the 1/$Q^4$ behavior, and this falloff 
appears to be fairly clear for $F^d_2$.  However, this behavior is sensitive
to the parameterization of $\gen$ at $Q^2$$>$2~(GeV/c)$^2$, which is
constrained directly only by the two data points from Ref.~\cite{riordan10}. 
Similarly, all of the calculations have the down-quark form factors falling
somewhat faster than 1/$Q^4$, suggesting that the apparent scaling behavior
may not continue to higher $Q^2$.  However, it is still clear that the
down-quark contributions fall significantly more rapidly than the up-quark
contributions at large $Q^2$, yielding a decrease in the magnitude of
$F_2^d/F_2^u$ at high $Q^2$. The very different $Q^2$ dependence for the up-
and down-quark contributions suggests that approximate 1/$Q^2$ scaling of
$F_2/F_1$ for the neutron is only approximate, and may be unrelated to the
predicted scaling behaviors.

The faster falloff of the down-quark contributions was interpreted in
Ref.~\cite{CJRW2011} and references therein as an indication of the
possibility of sizable nonzero strange matrix elements at large $Q^2$ or the
importance of diquark degrees of freedom.  While existing measurements of
parity-violating elastic scattering yield very small contributions from the
strange quarks up to $Q^2 \approx 1$~(GeV/c)$^2$~\cite{aniol04, armstrong05,
androic10, ahmed12}, they still leave open the possibility for significant
contributions from $G_E^s$ and $G_M^s$ which cancel in the parity-violating
observables~\cite{ahmed12, armstrong12}, although there are also results from
Lattice QCD that the strange-quark contribution is small for both the charge
and magnetic form factors~\cite{leinweber05, paschke11}. In the diquark model,
the singly-occurring down quark in the proton is more likely to be associated
with an axial-vector diquark than a scalar diquark, and the contributions of
the axial-vector diquark yields a more rapid falloff of the form factor.  The
up quarks are generally associated with the more tightly bound scalar diquarks,
yielding a harder form factor~\cite{roberts07,cloet09,wilson12}.

The flavor dependence of the nucleon form factors as obtained in the CJRW
extractions was reproduced quite well by incorporating the Regge contribution
into Generalized Parton Distributions (GPDs) that already apply diquark
models~\cite{hernandez2012}. Inclusion of the Regge contributions is crucial
to obtain the correct normalized structure functions. Therefore, the flavor
dependence was attributed mainly to Reggeon exchanges and quark-diquark
contributions. For the Dirac $Q^4 F^u_1$ form factor, the diquark contribution
dominates the Regge contribution at low $Q^2$ and both contributions become
comparable at high $Q^2$. On the other hand, for $Q^4 F^d_1$, both the diquark
and Regge contributions are rather comparable. For the Pauli $Q^4F^{(u,d)}_2$
form factors, the Regge contribution dominates that of the diquark contribution
and, in particular, at high $Q^2$.  This again shows the importance of the
diquark contribution at high $Q^2$, although in this framework, the Regge
contributions are important in achieving a better result at low $Q^2$.  These
data will also allow a flavor separation at higher $Q^2$ values, where the pion
cloud contributions, neglected or included in a less detailed fashion, are
expected to be smaller.  This will allow for a more direct test of the 
calculations of the three-quark core, with reduced uncertainties associated
with the more poorly understood pion cloud contributions.67

\begin{figure}[!htbp]
\begin{center}
\includegraphics*[width=7.9cm,height=5.5cm]{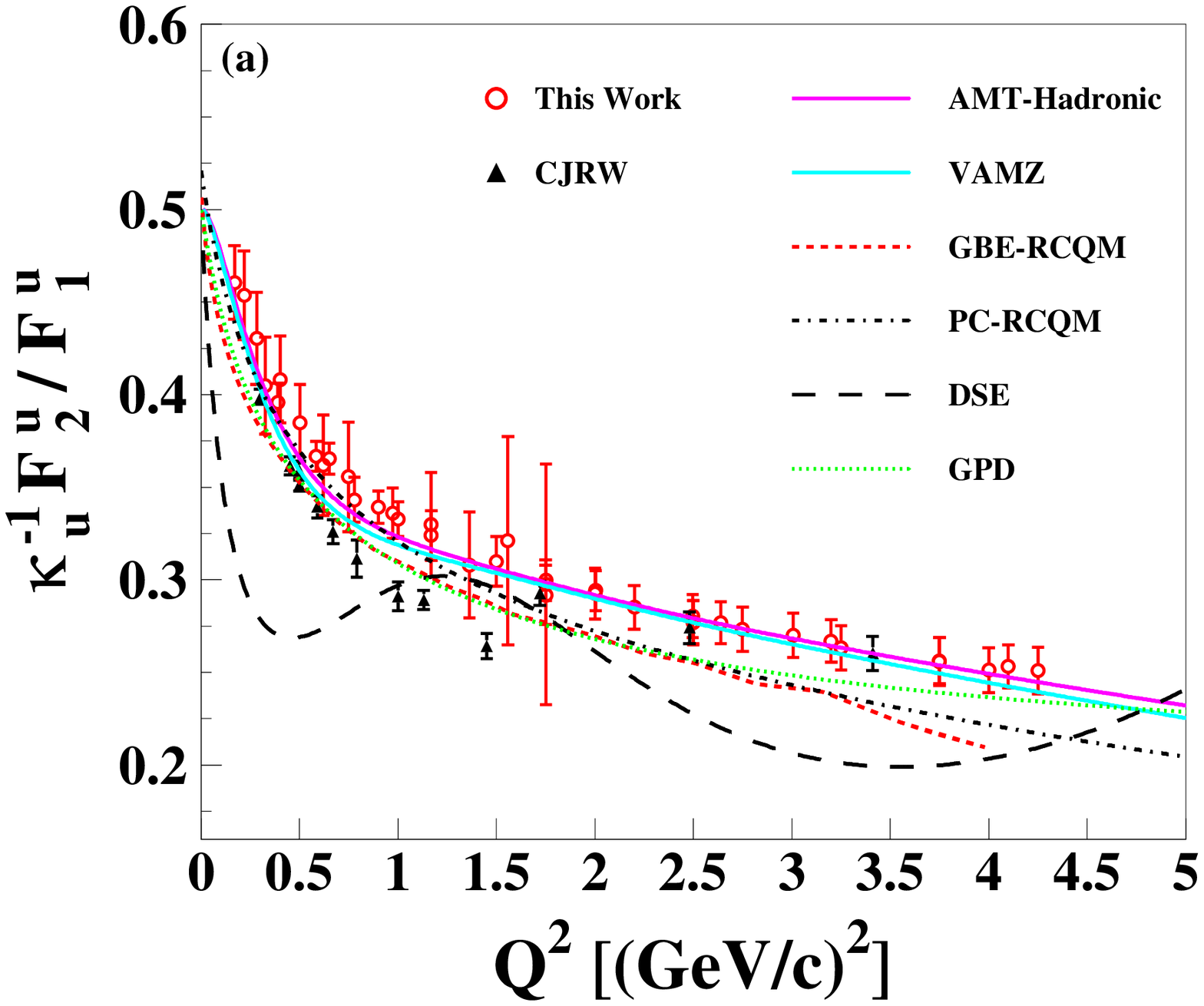}\\
\includegraphics*[width=7.9cm,height=5.5cm]{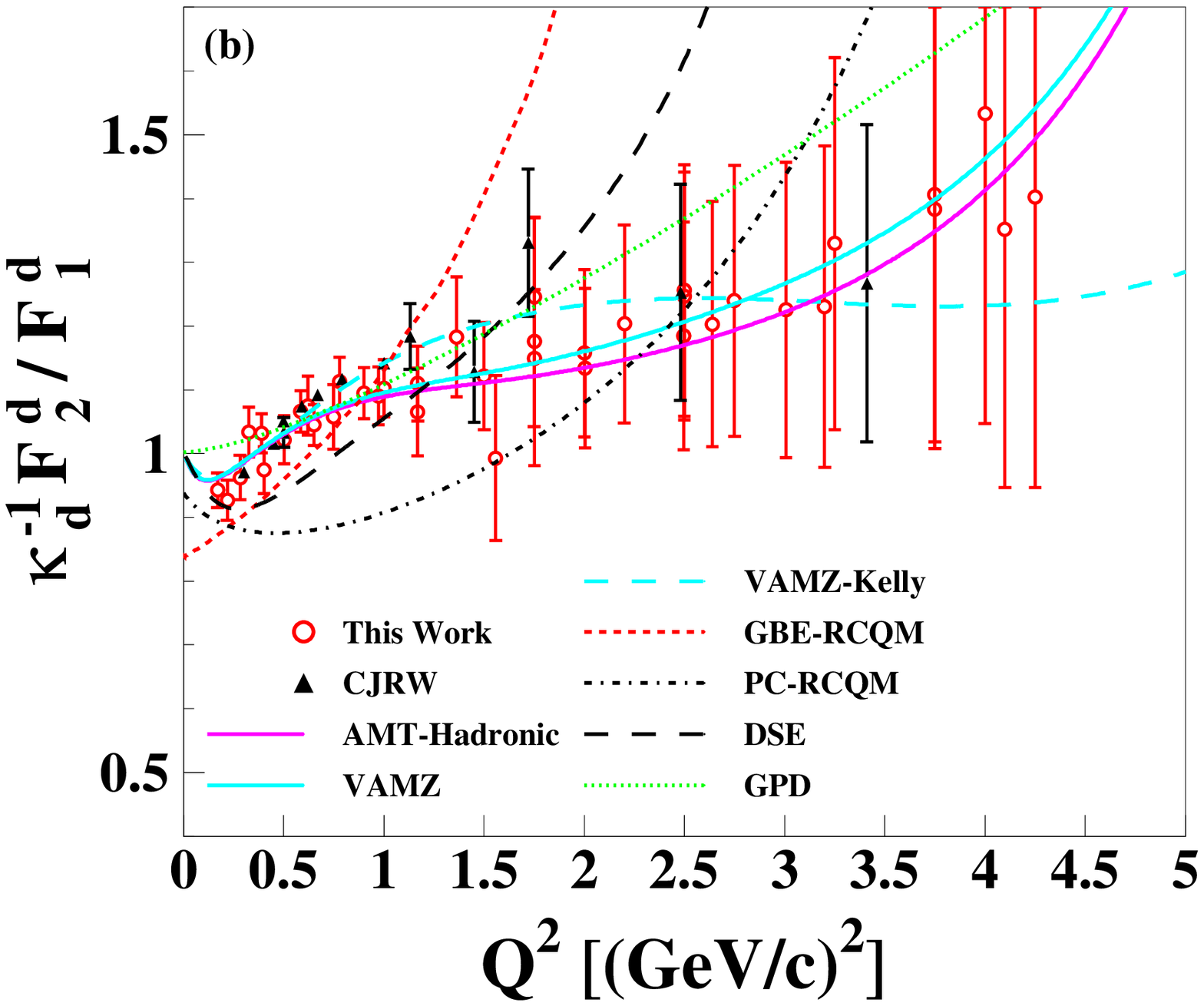}
\end{center}
\vspace{-0.5cm}
\caption{(color online) The ratios $\kappa^{-1}_u F^u_2/F^u_1$ [top], and
$\kappa^{-1}_d F^d_2/F^d_1$ [bottom].
}
\label{fig:F2u1uF2d1dRatios}
\end{figure}

The ratios $\kappa^{-1}_u F^u_2/F^u_1$ and $\kappa^{-1}_d F^d_2/F^d_1$ are
shown in Fig.~\ref{fig:F2u1uF2d1dRatios}. The values $\kappa_{u,d}$ 
are the $Q^2=0$ limits of $F_2^{u,d}$, $\kappa_u = \mu_u - 2 = 1.67$
and $\mu_d - 1 = -2.03$, where the fact that the magnetic form factor
contributions are normalized by quark charge but not by the number of valence
quarks yields the subtraction of 2 for the up quark contribution.  After
scaling by $\kappa_{u,d}^{-1}$, the ratios are normalized to $1/F_1^{u,d}$,
yielding 0.5 for the up-quark contribution and 1 for the down-quark.
$F^u_2/F^u_1$ falls rapidly at low $Q^2$, but the decrease is signficantly
slower above $Q^2=1$~(GeV/c)$^2$; note the offset zero in 
Fig.~\ref{fig:F2u1uF2d1dRatios}(a). Our values are somewhat larger than those
obtained in the CJRW extractions for $Q^2<$ 1.5 (GeV/c)$^2$, due to the
difference in the $F^u_2$ values. For the ratio $\kappa^{-1}_d F^d_2/F^d_1$,
the extractions from the data and the fits to the form factors yield
consistent results, with a slight increase in the ratio at low $Q^2$ and a
nearly constant value above 1~(GeV/c)$^2$. However, all of the calculations
show a very different behavior, with $F_1^d$ falling more rapidly than $F_2^d$
at large $Q^2$ values, leading to a rapid rise in the ratio.  This is 
clearly an area where the models should be examined more carefully, although
the present measurements of $\gen$ do not rule out a significant rise in the
ratio above 1.5~(GeV/c)$^2$. Measurements of $\gen$ planned for higher $Q^2$
after the Jefferson Lab 12 GeV upgrade~\cite{dudek12} will be critical in
pinning down the behavior of this ratio.

\section{conclusions} \label{conclusions}

We have extracted the flavor-separated contributions to the elastic nucleon
electromagnetic form factors based on parameterizations of the neutron form
factors and their uncertainties, and proton form factor extractions that
include phenomenological TPE corrections.  The extraction is similar to that
of the recent CJRW~\cite{CJRW2011} analysis, but with an explicit treatment of
two-photon exchange effects and the uncertainties on the proton form factor
and the neutron magnetic form factors.  The treatment of the TPE contributions
yields differences in some of the results at low $Q^2$ values, up to
$\approx$1.5~(GeV/c)$^2$.  In addition, while our updated parameterization of
$\gmn$ yields only a small change in the high-$Q^2$ behavior, it has a
significant impact near $Q^2 = 1$~(GeV/c)$^2$, where the recent CLAS
measurements~\cite{lachniet09} extraction is somewhat below earlier
extractions~\cite{anklin98,kubon02}.  As our updated fit falls in between
these measurements, the difference between our parameterization and the 
Kelly fit represents a reasonable estimate of the uncertainty, given these
inconsistent extractions.  A new extraction of $\gmn$ in this region 
will be important to help resolve this issue.
The additional uncertainties included in this analysis generally have a small
impact, but are more important for some of the flavor-separated form factors
and quantities which are insenstivie to $\gen$. The extracted flavor-separated
form factors are qualitatively reproduced by a range of models, with better
quantitative agreement for those models which constrain parameters by directly
fitting to nucleon form factor data.

The strong linear falloff with $Q^2$ observed in $G_E^p/G_M^p$ is not present
in either the up- or down-quark contributions, but arises due to a
cancellation between a weaker $Q^2$ dependence for the up quark and a negative
but relatively $Q^2$-independent contribution from the down quark. This
indicates that the rapid falloff and the zero crossing expected near
$Q^2=10$~(GeV/c)$^2$ is associated more with the difference between the up- and
down-quark distributions than by the isospin-averaged spatial density
distributions.

As noted in the previous analysis~\cite{CJRW2011}, the $F_1$ and $F_2$ form
factors show a different $Q^2$ dependence for up- and down-quark
contributions, which can be shown to be a consequence of diquark degrees of
freedom in several of the calculations.  We see indications that the
down-quark contributions to the Dirac and Pauli form factors deviate from the
1/$Q^4$ scaling suggested in the CJRW analysis, and observe that there are
small differences between the $Q^2$ dependence in $F_1$ and $F_2$ for both the
up- and down-quark contributions.

Finally, the up and down quarks yield very different contributions to
$G_E/G_M$ (and $F_2/F_1$), with the ratio decreasing slowly with $Q^2$ for the
up quark and increasing rapidly for the down quark.  The down-quark
contribution to this ratio is not well reproduced by any of the calculations,
and even the qualitative behavior is not reproduced in most approaches.
Data at higher $Q^2$ will better constrain the behavior of these ratios,
and allow for a more detailed evalation of the nucleon models in a region
where pion cloud contributions are expected to be less significant.

\begin{acknowledgments}

This work was supported by Khalifa University of Science, Technology and
Research and by the U.~S. Department of Energy, Office of Nuclear Physics,
under contract DE-AC02-06CH11357. We are grateful to I. C. Clo{\"e}t,
J. O. Gonzales-Hernandez, S. Liuti, G. Miller, W. Plessas and C. D. Roberts for
providing us with their calculations. \end{acknowledgments}

\bibliography{flavordecomp}

\end{document}